\documentclass[aps,prb,twocolumn,amsmath,superscriptaddress,amssymb,longbibliography]{revtex4}
\usepackage{graphicx}
\usepackage{enumitem}
\usepackage[colorlinks,bookmarks=false,citecolor=red,linkcolor=blue,urlcolor=blue]{hyperref}

\newcommand{\bea}{\begin{eqnarray}}          
\newcommand{\eea}{\end{eqnarray}}          
\newcommand{\la}{\langle}          
\newcommand{\ra}{\rangle}

\allowdisplaybreaks

\begin{document}

\title{Bound states without potentials: localization at singularities}
\author{Eric He}
\email{heeric@berkeley.edu}
\affiliation{University of California, Berkeley, California 94720, USA}
\affiliation{Department of Physics, Brock University, St. Catharines, Ontario L2S 3A1, Canada}
\author{R. Ganesh}
\email{r.ganesh@brocku.ca}
\affiliation{Department of Physics, Brock University, St. Catharines, Ontario L2S 3A1, Canada}
\date{\today}

\begin{abstract}
Bound state formation is a classic feature of quantum mechanics, where a particle localizes in the vicinity of an attractive potential. This is typically understood as the particle lowering its potential energy. 
In this article, we discuss a paradigm where bound states arise purely due to kinetic energy considerations. This phenomenon occurs in certain non-manifold spaces that consist of multiple smooth surfaces that intersect one another. The intersection region can be viewed as a singularity where dimensionality is not defined. 
We demonstrate this idea in a setting where a particle moves on $M$ spaces ($M=2, 3, 4, \ldots$), each of dimensionality $D$ ($D=1, 2$ and $3$). The spaces intersect at a common point, which serves as a singularity. To study quantum behaviour in this setting, we discretize space and adopt a tight-binding approach. We generically find a ground state that is localized around the singular point, bound by the kinetic energy of `shuttling' among the $M$ surfaces. We draw a quantitative analogy between singularities on the one hand and local attractive potentials on the other. To each singularity, we assign an equivalent potential that produces the same bound state wavefunction and binding energy. The degree of a singularity ($M$, the number of intersecting surfaces) determines the strength of the equivalent potential.
With $D=1$ and $D=2$, we show that any singularity creates a bound state. This is analogous to the well known fact that any attractive potential creates a bound state in 1D and 2D. 
In contrast, with $D=3$, bound states only appear when the degree of the singularity exceeds a threshold value. This is analogous to the fact that in three dimensions, a threshold potential strength is required for bound state formation. We discuss implications for experiments and theoretical studies in various domains of quantum physics.
\end{abstract} 
                                 
\maketitle
\section{Introduction}
A free particle, in its quantum mechanical ground state, typically spreads uniformly to occupy all available space. This allows the particle to lower its kinetic energy. However, in the presence of an attractive potential, it may localize into a bound state to lower its potential energy. This phenomenon reflects competition between kinetic and potential energies. In this article, we discuss a paradigm where bound states form without any potentials. Rather, the particle moves on a singular space consisting of multiple surfaces that intersect at a `junction'. This allows for a new type of kinetic energy that favours localization of the particle. We present a detailed analysis of this phenomenon, focussing on the role of dimensionality.

\begin{figure}
\includegraphics[width=3.3in]{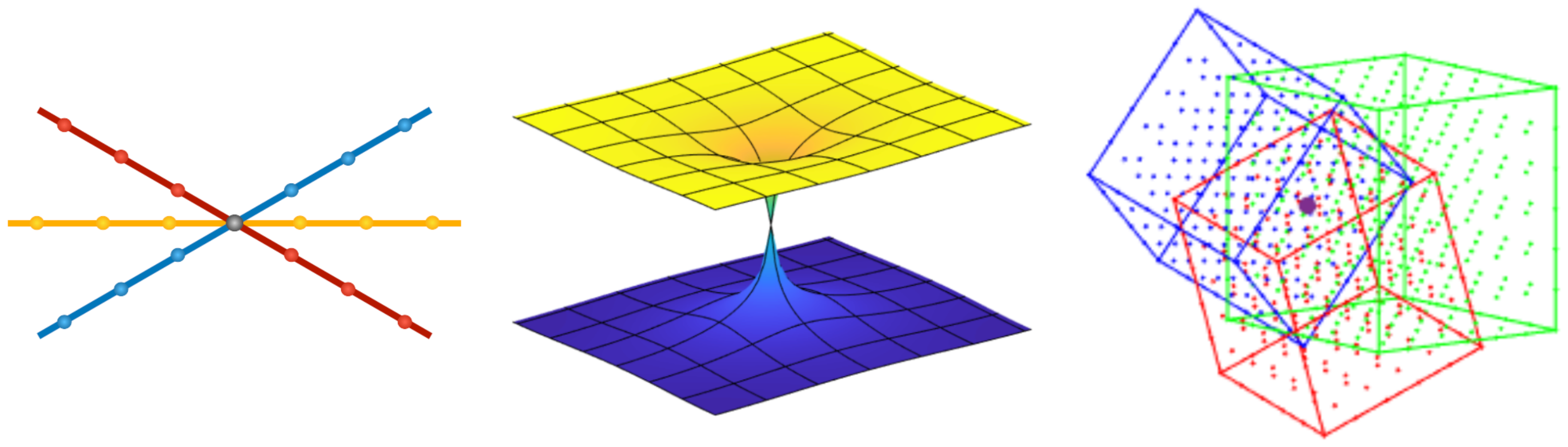}
\caption{Self-intersecting spaces. Left: Three wires (one-dimensional spaces) that intersect at a point. Centre: Two sheets (two-dimensional spaces) that intersect at a point. Right: Three cubes (three-dimensional spaces) that intersect at a point.  } 
\label{fig.spaces}
\end{figure}

Quantum mechanics on intersecting spaces is a concrete, testable proposition. Experiments with semiconductor architectures have explored X-junctions and T-junctions. Indeed, bound states have been seen where electrons/holes are localized near junctions\cite{Sols1989,Berggren1991,Berggren1992,Exner1996}. Collective excitonic excitations have also been found to bind to junctions\cite{Schult1989,Gaididei_1992,Goni1992,Hasen1997}. Analogous phenomena may also occur in classical wave mechanics, e.g., in junctions of photonic\cite{Bulgakov2002} or phononic waveguides\cite{Maksimov2006,Nakarmi2021}. 
Recently, singular spaces have been invoked to describe quantum magnets. At low energies, the physics of a quantum magnet resembles that of a particle moving on the space of classical ground states\cite{Khatua2019,khatua_thesis}. In certain magnetic clusters, frustration leads to complex spaces that contain intersecting wires/sheets. At very low energies, the magnet freezes at the intersection, i.e., it orders in a particular classical configuration. This is equivalent to a particle binding to a junction due to bound state formation. This effect has been called `order by singularity', as a special case of the well known order-by-disorder phenomenon\cite{Khatua2019,Sarvesh2020,Khatua2021,khatua_thesis}. In these various experimental settings, junctions induce bound states with dramatic physical consequences. The goal of this article is to develop an understanding of this phenomenon, its physical origin and organizing principles.

We consider a class of spaces consisting of multiple surfaces that intersect one another. The locus of intersection represents a singularity. In the discussion below, we characterize each singularity by two quantities: co-dimension and degree. Fig.~\ref{fig.spaces} shows some examples. On the left, we have three one-dimensional channels that intersect at a point-like junction. A generic point in this space lies on one of the channels; its local neighbourhood is one-dimensional. In contrast, the junction is a zero-dimensional region (a point) whose local neighbourhood does not have a well-defined dimensionality. 
The difference between these two dimensionalities is the co-dimension, which is unity here. 
In addition, we assign a degree of three -- representing the number of channels that meet at this junction. In Fig.~\ref{fig.spaces} (centre), we see two sheets that intersect at a point. This represents a singularity of co-dimension two. The space is generically two-dimensional, while the singularity is point-like. As we have two sheets, the degree is two. 
In Fig.~\ref{fig.spaces} (right), we see three cubes which are understood to share a common point. The common point is a zero-dimensional singularity. The co-dimension and the degree are three in this case.

For comparison, we will also discuss bound states induced by attractive potentials on smooth spaces. There is extensive literature available on bound states induced by various potentials. It is well known that dimensionality plays a key role. In 1D and 2D, an infinitesimal potential suffices to create a bound state. However, in three and higher dimensions, a threshold potential strength is required\cite{Simon1976,Yang1989,Nieto2002,Geltman2011}. In this article, we use a tight binding approach that can handle both singularities and potentials on the same footing. Naively, the problem of a potential appears to be very different from that of a singularity. However, our results bring out a deep connection. As we show below, any singularity is quantitatively equivalent to a potential, in the sense that it produces the same bound state.

\section{The tight-binding approach}
The traditional approach to quantum mechanics is to construct a Hamiltonian operator and find its eigenfunctions. This cannot be carried out on singular spaces as the Hamiltonian cannot be written down in the vicinity of a singularity. For example, a gradient operator cannot be defined near a junction of two wires. Solutions can still be found using ad hoc methods. For example, eigenfunctions can be found on each smooth segment, with a suitable boundary condition imposed at the singularity. The choice of boundary condition can affect the result\cite{Gratus1994,Kostrykin1999,Andrade2016}. 

In this article, we take an alternative approach using tight-binding. 
Originally developed to describe band structure in solids, tight binding typically describes an electron in a lattice of atoms\cite{Ashcroft_book,Allison_book,Harrison2004}. The Hilbert space is spanned by localized wavefunctions centred on each atom. If the lattice constant is not too small, an electron hops from an atom to any of its neighbours. Here, we adapt this approach with `atoms' arranged on a singular space rather than forming a regular lattice.

The tight-binding approach lends itself to an ambiguity-free procedure for a one-dimensional problem, such as the one shown in Fig.~\ref{fig.spaces} (left). In two dimensions and higher, there can be multiple ways of discretizing a smooth surface. For example, a smooth two-dimensional region can be discretized into a square or a triangular grid. Once this choice is made, there is no further ambiguity in the procedure or the solutions obtained. In our calculations, we choose a square (cubic) discretization for two (three) dimensions.

In the rest of this article, we solve free-particle tight-binding Hamiltonians of the form
\bea
H = -t \sum_{\langle mn \rangle} \big\{  c_{m}^\dagger c_n + c_{n}^\dagger c_m\big\} ,
\label{eq.Hgen}
\eea
where the hopping amplitude, $t$, sets the energy scale. The sum runs over nearest-neighbour bonds, with $m$ and $n$ representing sites at the end of each bond. The operator $c_{m}^\dagger$ creates a particle at site $m$, while $c_n$ annihilates a particle at site $n$. The Hamiltonian can be viewed as encoding time evolution on a discrete graph. A particle that is initially localized at one site can hop to the immediate neighbours in one step. Upon repeated action of the Hamiltonian, the particle may hope onto the next-nearest neighbours and further.

The geometry of the space is encoded in the assignment of neighbours.
Away from the singularity, each site has $2D$ neighbours, where $D$ is the dimensionality of the surface. As shown in Fig.~\ref{fig.spaces}, the singularity is a single site with $(2M D)$ neighbours, where $M$ is the degree of the singularity; we have $2D$ neighbours per surface, with $M$ surfaces in total. Operationally, we take each surface to have linear dimension $L$ with periodic boundaries. The total number of sites in the problem is then $N=ML^D-M+1$. The resulting Hamiltonian is an $N \times N$ symmetric matrix. Its eigenvectors represent stationary states, while the eigenvalues yield the corresponding energies. For small system sizes, we carry out full diagonalization to find all eigenvectors and eigenvalues. For large systems, we take advantage of the sparse character of the tight binding Hamiltonian and employ Krylov-space-based routines to find the lowest few eigenstates. 

In the tight binding setup, an eigenstate satisfies the following relation at every site:
\bea
-t \sum_{m(n)} \psi_m = E \psi_n,
\label{eq.cond}
\eea
where $n$ represents any given site. The index $m(n)$ runs over the neighbours of $n$, $\psi_p$ represents the eigenvector component at site $p$ and $E$ represents the eigenvalue.

A bound state can be identified in two ways: from the eigenvector or from the eigenvalue. The eigenvector must be peaked at the singularity, decaying to zero as we move away. The eigenvalue must lie below a threshold value, $-2Dt $, representing the lowest value possible for a delocalized state on a $D$-dimensional surface. This can be expressed in terms of a binding energy, $E_{binding} = -2Dt - E_{state}$. A bound state must have a positive binding energy. The higher the binding energy, more bound is the state.

For the one-dimensional tight binding problem with a setup as shown in Fig.~\ref{fig.spaces} (left), bound state(s) can be found using analytic arguments. They are exponentially localized around the singularity, as we show below. More generally, the eigenvalues and eigenvectors can be found numerically.

\section{$D=1$: Intersecting wires}

\subsection{Analytic solution to the tight-binding problem}

We first discuss the case of $M$ one-dimensional wires intersecting at a point-like junction. The arguments in this section were first presented in Ref.~\onlinecite{Khatua2021} in the context of a certain magnetic model. To represent the wavefunction, we denote the junction site as $j=0$. To all other sites, we assign an integer value that encodes distance from the junction. For example, the immediate neighbours of the junction are assigned $j=1$. 
This tight binding problem produces one bound eigenstate, described by the ansatz 
\bea
\psi_{j} = \frac{1}{\mathcal{N}} e^{-\alpha_M j};~~~\psi_{0} = \frac{1}{\mathcal{N}}.
\label{eq.psi1dsing}
\eea  
Here, $\mathcal{N}$ is a normalization constant. We demand that this be an eigenstate with eigenvalue $E_M$. At a site away from the junction, the eigenstate condition of Eq.~\ref{eq.cond} yields
 \bea
 -t (e^{-\alpha_M}+e^{\alpha_M}) = E_M.
 \label{eq.genericsite}
 \eea
 At the junction site, the same condition yields
 \bea
 -2Mt e^{-\alpha_M} = E_M.
 \label{eq.junction}
 \eea
 From Eqs.~\ref{eq.genericsite} and \ref{eq.junction}, we solve for $\alpha_M$ and $E_M$,
\bea
\alpha_M = \frac{1}{2} \ln \{2M-1\},~~E_M=-2Mt/\sqrt{2M-1}.
\label{eq.bsM}
\eea
Note that for $M=1$, there is no singularity as we only have one wire. In this limit, the bound state vanishes as $\alpha_{M=1}=0$ and $E_{M=1} = -2t$. For $M\geq2$, $\alpha_M$ represents a decay constant. The binding energy is given by $E_{binding,D=1} = (-2t - E_M)$. The normalization constant can be explicitly found, $\mathcal{N} = \sqrt{\frac{2M-1}{M-1}}$.

For $M\geq2$, $\alpha_M$ monotonically increases with $M$. In parallel, $E_M$ monotonically decreases or equivalently, the binding energy monotonically increases. This shows that the state becomes progressively more bound as the degree of the singularity increases. For very large $M$, the bound state is entirely localized at the singularity. 

\subsection{Comparison with bound states induced by a potential}

The bound state at the singularity can be compared to one induced by a local attractive potential. Within the tight-binding approach, we consider a smooth one-dimensional chain with sites labelled by $j$, a coordinate that runs over all integers. At $j=0$, we have an on-site attractive potential of strength $g$. The Hamiltonian is given by 
\bea
H_{1D,g} = -t \sum_{j} \big\{ c_{j}^\dagger c_{j+1} + c_{j+1}^\dagger c_{j}  \big\} -g c_{0}^\dagger c_{0}.
\label{eq.Hpot}
\eea
This problem also generates a bound state, with the wavefunction  
\bea
\psi_{j} = \frac{1}{\mathcal{N}'}e^{-\alpha_g \vert j \vert}.
\label{eq.psi1dpot}
\eea
where $\mathcal{N}'$ is a normalization constant. For $j \neq 0$, the eigenstate condition takes the same form as Eq.~\ref{eq.genericsite}. At $j=0$, the condition is modified by the potential to give
\bea
-2 t e^{-\alpha_g} - g = E_g,  
\eea
where $E_g$ is the eigenvalue. 
It is convenient to express the potential strength and the energy eigenvalue as dimensionless quantities, using $\tilde{g} \equiv g/2t$ and $\tilde{E}_g = E_g/2t$. In terms of these quantities, we find
\bea
\alpha_g = \ln \{ \tilde{g} + \sqrt{\tilde{g}^2 + 1} \};~~\tilde{E}_g = \frac{-\tilde{g}^2 - \tilde{g} \sqrt{\tilde{g}^2 + 1}-1}{\tilde{g} + \sqrt{\tilde{g}^2 + 1}}.
\label{eq.bsg}
\eea
The normalization constant comes out to be $\mathcal{N}' = \sqrt{1+\frac{1}{\tilde{g}\{\tilde{g} + \sqrt{\tilde{g}^2 + 1}\}}}$. These values describe a bound state induced by a potential on a smooth one-dimensional space with no singularities. In contrast, those in Eq.~\ref{eq.bsM} describe a bound state created at a singularity of co-dimension 1 and degree $M$, with no potential involved. Remarkably, the wavefunctions have the same form in both cases as given by Eqs.~\ref{eq.psi1dsing} and \ref{eq.psi1dpot}. This allows us to draw a precise equivalence, $M \leftrightarrow \tilde{g}_M$, where $\tilde{g}_M$ satisfies
\bea
M = \Big[\tilde{g}_M^2 +  \tilde{g}_M\sqrt{\tilde{g}_M^2+1}+1\Big].
\eea   
The equivalence can be stated as follows. On the one hand, we consider a singularity of co-dimension 1 and degree $M$, with no potential. On the other hand, we consider a potential of strength $\tilde{g}_M$ on a smooth one-dimensional chain. These two situations produce bound states with precisely the same decay constant \textit{and} binding energy.

The equivalent potential, $\tilde{g}_M$, increases monotonically with $M$. For large $M$, we see that $\tilde{g}_M \sim \sqrt{M}$.

\subsection{Mechanism for binding}

A bound state has lower energy than the continuum of delocalized states. The underlying mechanism provides some way for the bound state to lower its energy. What is the mechanism in the case of a potential or in the case of a singularity? This question can be directly addressed within the tight binding approach where the Hamiltonian is a sum of local terms. We have one term for each bond, representing the kinetic energy of hopping between two sites. In the case of a potential, we also have an on-site potential energy. Given the ground state wavefunction, we may evaluate the contribution of each term to its energy. 

For the case of a potential-induced bound state, the Hamiltonian is given by Eq.~\ref{eq.Hpot} with the wavefunction given by Eq.~\ref{eq.psi1dpot}. Bound state formation is driven by the potential energy term. This can be seen by examining the potential energy contribution as a fraction of the bound state's energy, 
\bea
\frac{\la -g c_0^\dagger c_0 \ra}{{E}_g}  = \left(
\frac{\tilde{g}^2 + \tilde{g}\sqrt{\tilde{g}^2+1}}{\tilde{g}^2 + \tilde{g}\sqrt{\tilde{g}^2+1}+1}
\right)^2
= \left(\frac{-\tilde{g}}{\tilde{E}_g}\right)^2.
\eea
For very small $\tilde{g}$, this quantity approaches zero -- the state is well spread with low weight at the potential. At $\tilde{g} =1$, it becomes half. For large $\tilde{g}$, it approaches unity; the energy of the bound state comes almost entirely from the potential energy term. This reveals that the mechanism for bound state formation is potential-energy-lowering.

We now compare with the case of a bound state at a singularity. The Hamiltonian can be written in the form of Eq.~\ref{eq.Hgen} with the wavefunction given by Eq.~\ref{eq.psi1dsing}. Binding is driven by the $M$ bonds that connect outward from the singularity. We evaluate 
\bea
\frac{-t \sum_{j(0)}  \langle \big\{ c_{j}^\dagger c_0 + c_0^\dagger c_j \big\} \rangle }{E_M} = \frac{ 2M-2 }{2M-1}. 
\eea 
The index $j(0)$ runs over the 2M sites that are directly connected to the singularity. When $M=1$, this ratio vanishes as we have a delocalized ground state. For $M=2$, this ratio yields $2/3$, i.e., two-thirds of the ground state energy arises from the immediate vicinity of the singularity.
For large $M$, this ratio approaches unity. That is, all of the ground state's energy comes from the $2M$ bonds that are connected to the junction. This represents a kinetic energy contribution, arising from the particle's `shuttling' motion from one wire to another. This is a new form of kinetic energy that is not present on smooth surfaces. The ground state is bound due to shuttling-kinetic-energy that can only be gained in the vicinity of the singularity.

\section{$D=2$: Intersecting sheets} 

We next consider singularities of co-dimension 2, with an example shown in Fig.~\ref{fig.spaces} (centre). We consider the space of $M$ sheets that share a common point. We discretize the space using a square grid for each sheet, assuming that they intersect at the origin. This setup generates a single bound state for any $M\geq 2$. However, the wavefunction cannot be expressed as a simple analytic form. Instead, we present numerical solutions and fit them to a suitable functional form.

\subsection{Numerical solution to the tight-binding problem}

\begin{figure*}
\includegraphics[width=2.3in]{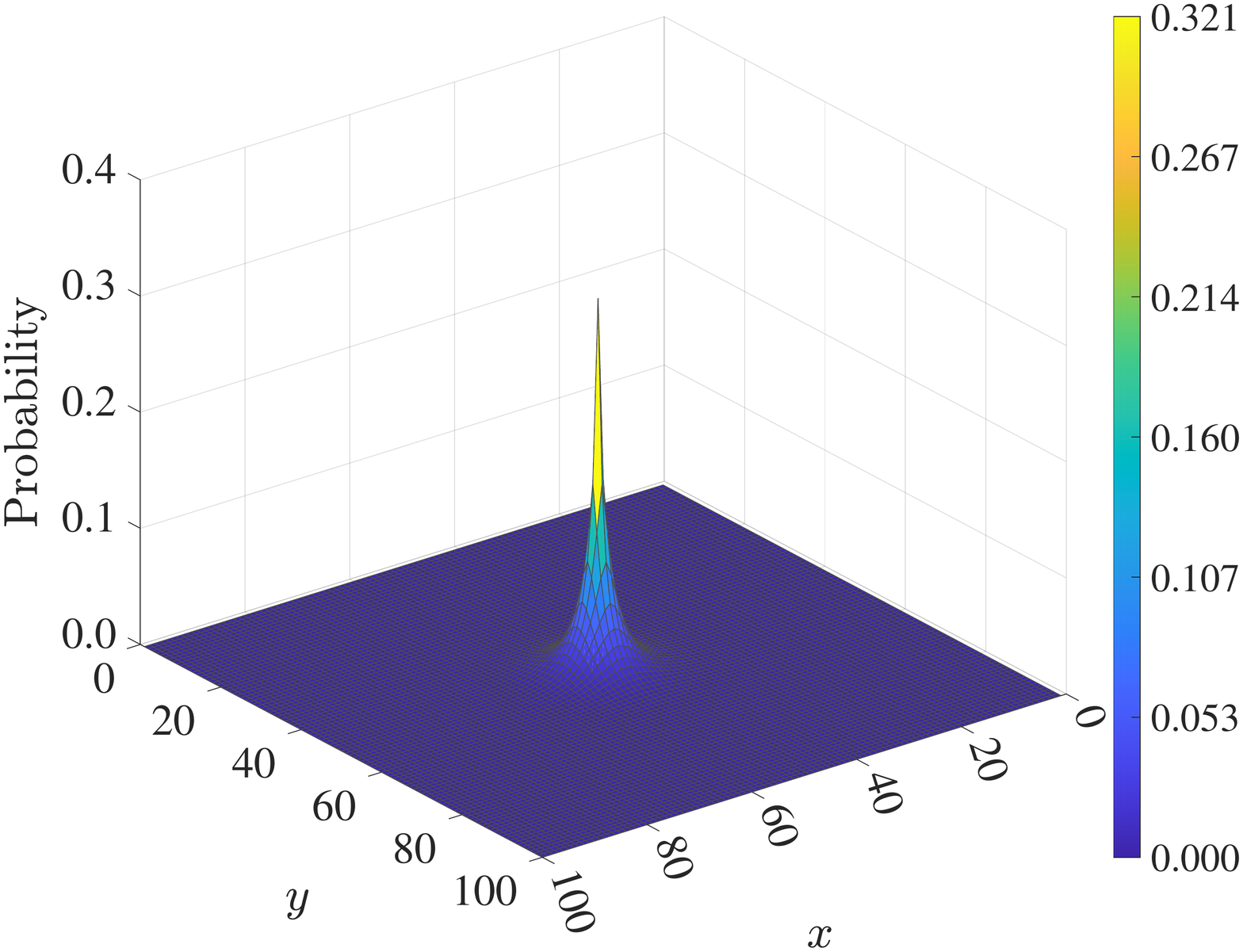}
\includegraphics[width=2.3in]{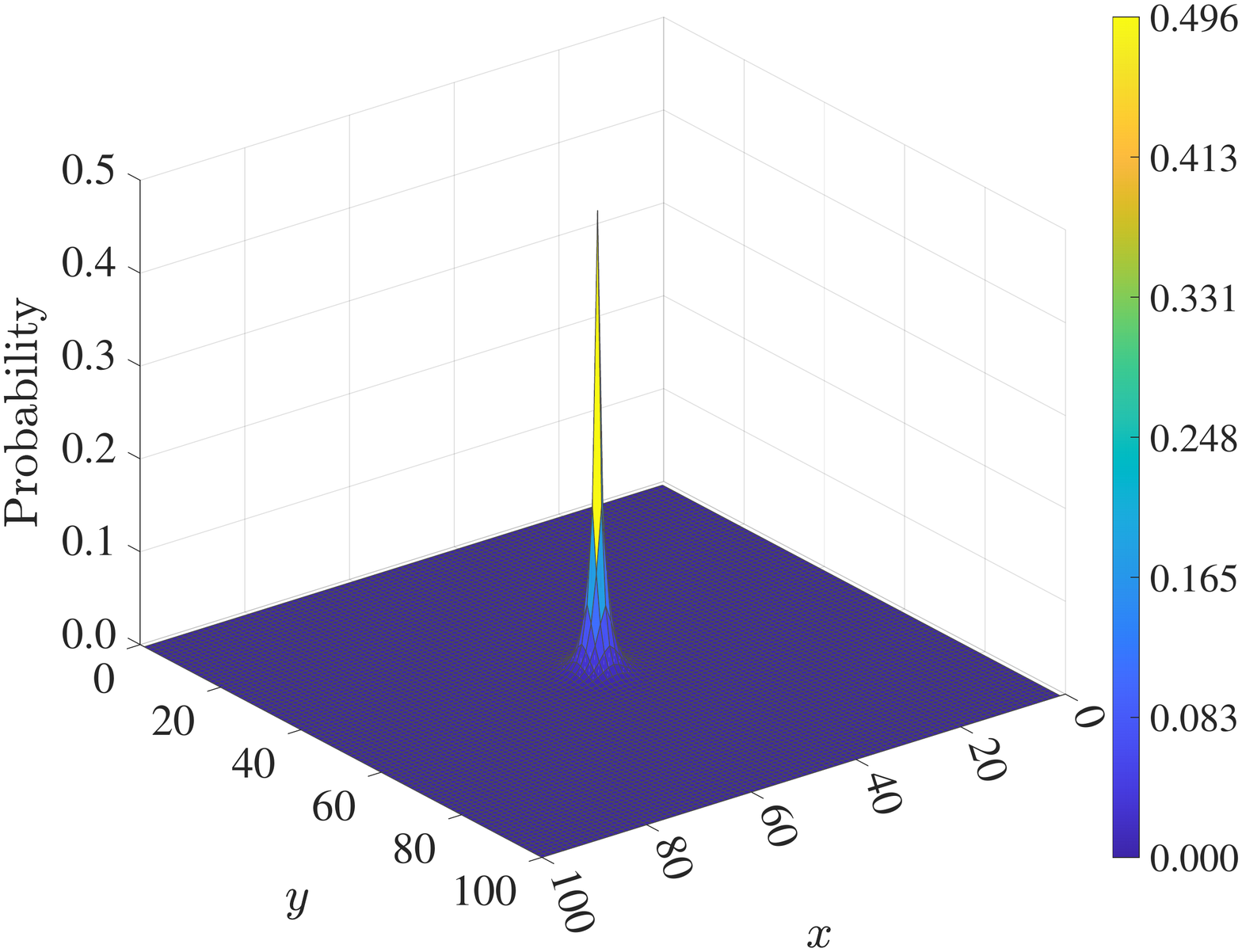} 
\includegraphics[width=2.3in]{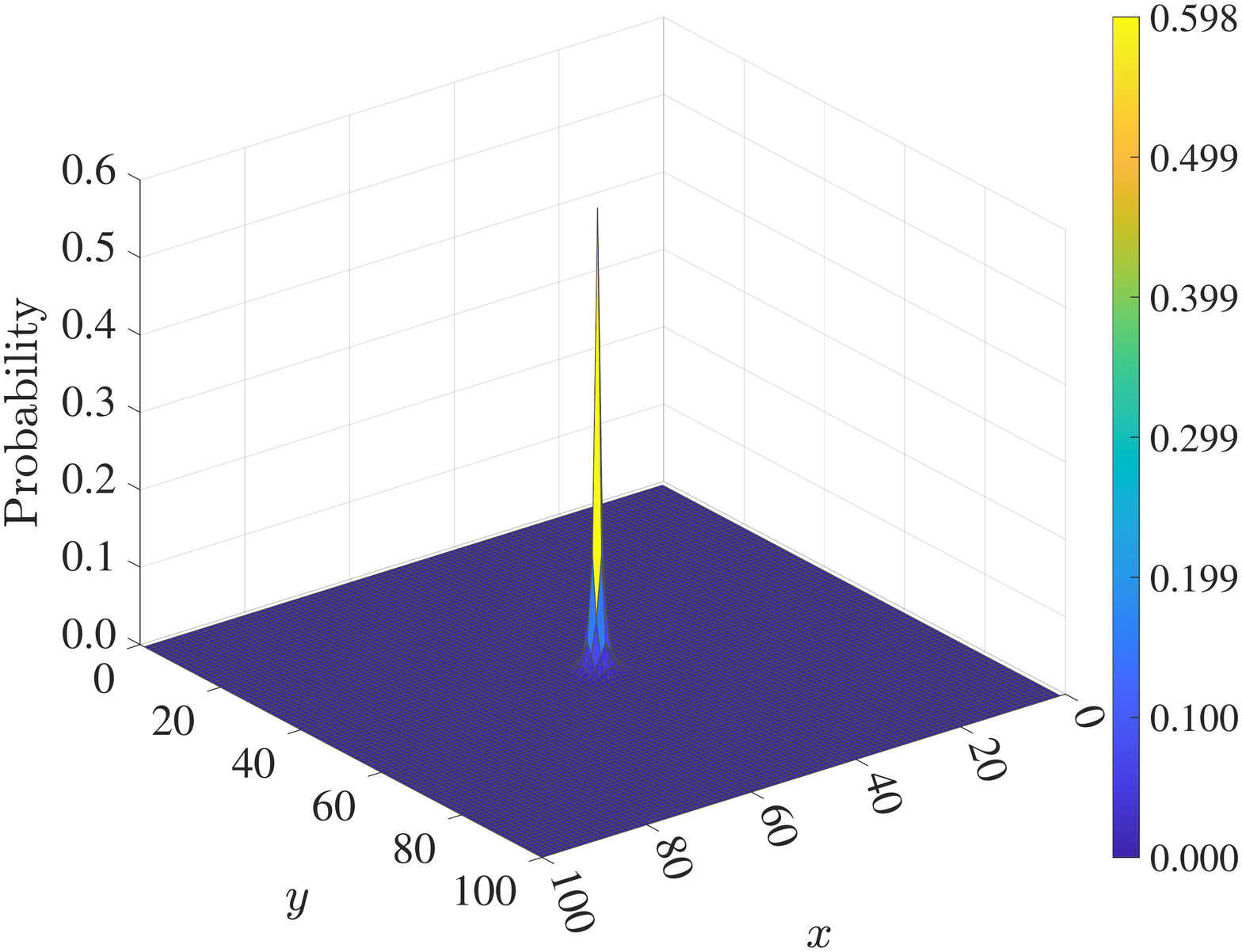}
\includegraphics[width=2.3in]{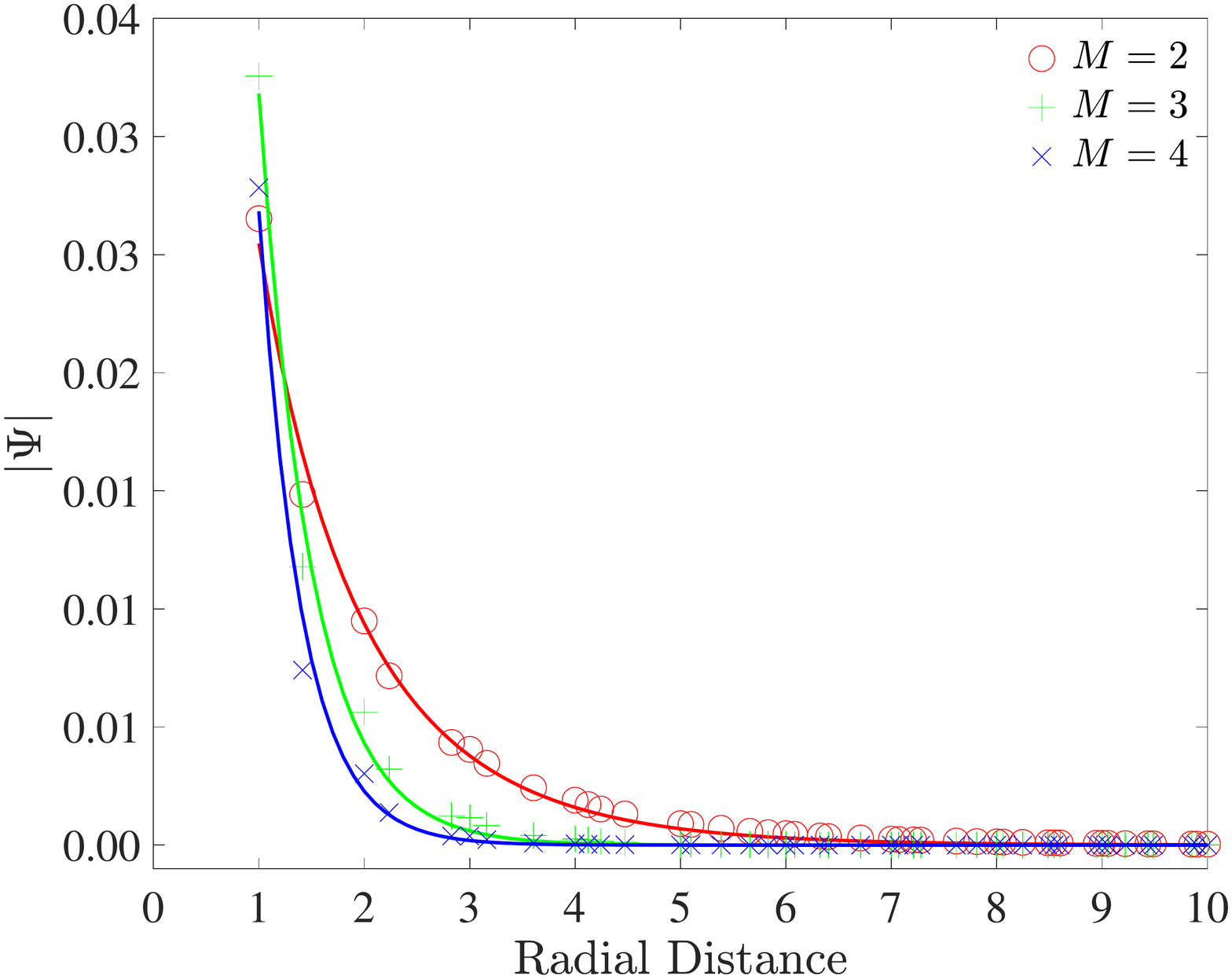}
\includegraphics[width=2.3in]{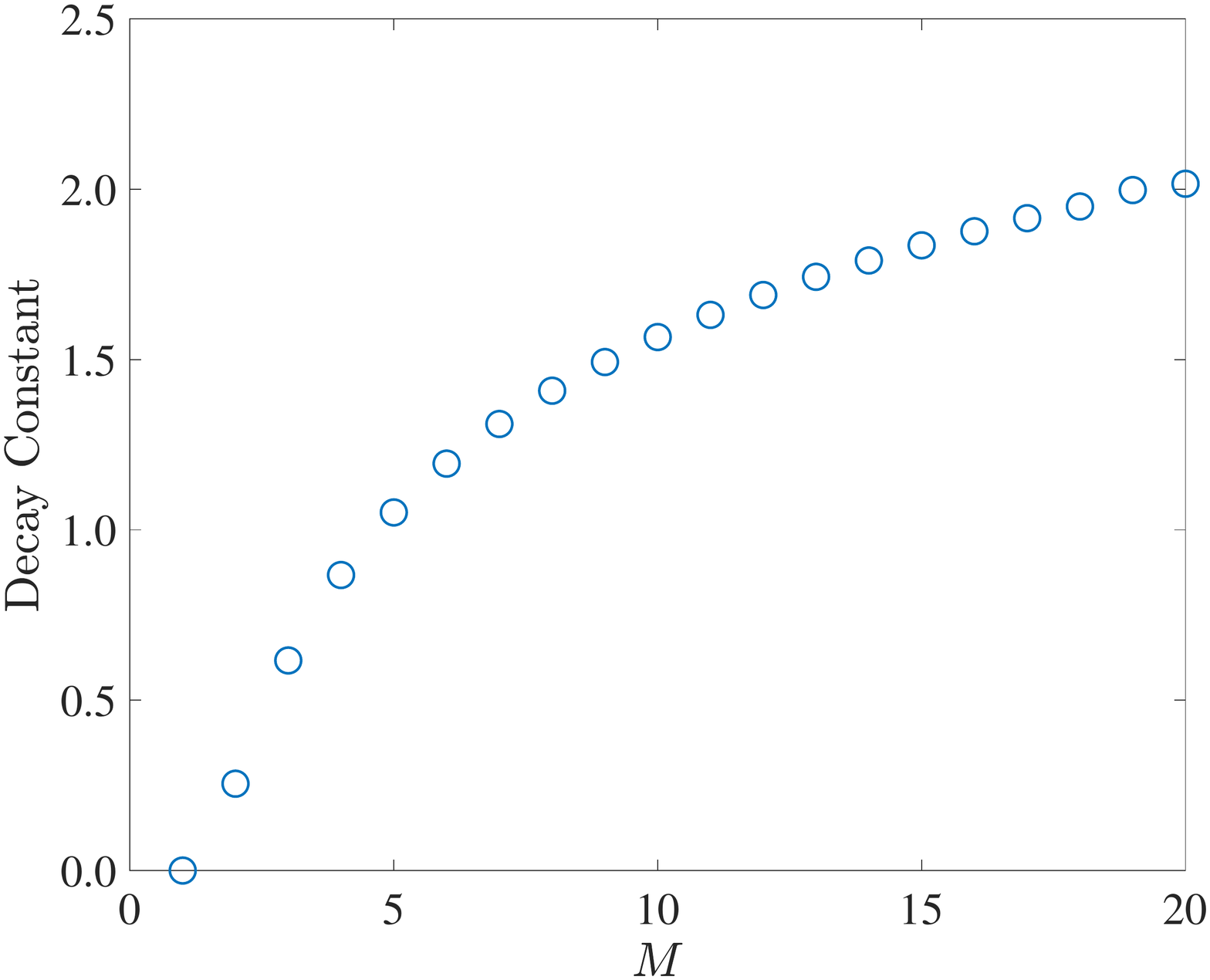}
\includegraphics[width=2.3in]{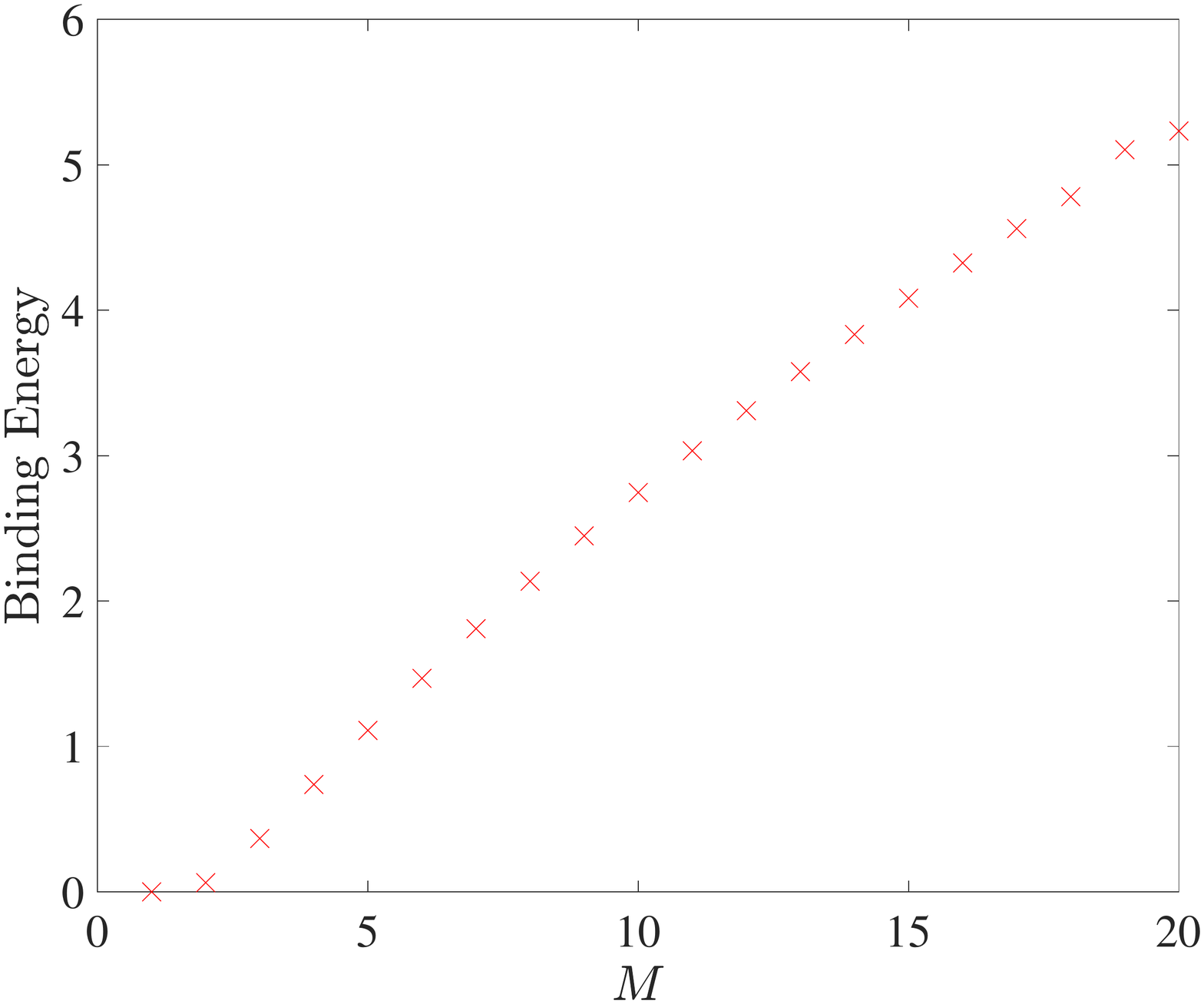}
\caption{The ground state on a space with $M$ 2D sheets intersecting at a point. The plots correspond to $M=2$ (top left), $M=3$ (top centre) and $M=5$ (top right). Each 2D sheet is taken to be a 100 $\times$ 100 grid with periodic boundaries. The origin is common to all the sheets. The plots show the spatial variation in the probability (amplitude squared). These wavefunctions fit well to a modified Bessel form of order 0 (bottom left). As the degree of the singularity increases, the decay constant obtained from the fit increases (bottom centre) as does the binding energy (bottom right). The binding energy is shown in units of $t$. } 
\label{fig.2d_sing_bss}
\end{figure*}

We consider $M$ sheets, each modelled as an $L\times L$ square grid. The sheets may have open or periodic boundaries. A generic point in this system has four nearest neighbours, as it lies on a two-dimensional sheet. In contrast, the central point in every sheet is taken to be the same. This point has $4M$ neighbours -- four on each of the $M$ sheets. 
This configuration defines a graph with $\{ML^2 - (M-1)\}$ sites. We solve the tight binding Hamiltonian of Eq.~\ref{eq.Hgen} on this graph by numerical diagonalization. We identify the lowest eigenstate and examine its wavefunction. As we show below, this state decays rapidly as we move away from the centre point. As it decays before reaching the boundaries, it is not sensitive to boundary conditions.

Fig.~\ref{fig.2d_sing_bss} (top row) shows examples of bound state wavefunctions. It depicts the probability (squared amplitude) of finding the particle at each site. The panels, from left to right, correspond to $M=2$, $3$ and $5$. In each case, the wavefunction on one of the intersecting sheets is shown -- the same wavefunction appears on every sheet. From the plots, we immediately see  localized character, with probability peaked at the singularity and decaying as we move away. With increasing degree of the singularity (increasing $M$), the ground state becomes more tightly bound.

The same information is shown as a two-dimensional plot in Fig.~\ref{fig.2d_sing_bss} (bottom left), with the wavefunction amplitude plotted against the radial coordinate (distance from the singularity). The solution has circular symmetry: sites with the same radial coordinate have the same wavefunction amplitude. Note that the phase is uniform at all points. As shown in Fig.~\ref{fig.2d_sing_bss} (bottom left), the amplitude is well fit by a function of the form 
\bea
f(r)= a~k_{0}(\gamma r + b).
\label{eq.modBesform}
\eea
Here, $k_{\nu}$ is the modified Bessel function of the second kind, of order $\nu$. This form is known from the continuum problem of a bound state induced by an attractive local potential in two dimensions, e.g., a square well potential (see Ref.~\onlinecite{Nieto2002, Geltman2011}). The bound wavefunction takes this form in the external region (outside the well). For each value of $M$, we obtain $a$, $b$ and $\gamma$ as fitting parameters. The coefficient $b$ represents a horizontal shift. For any $M\geq 2$, the best-fit value of $b$ is less than the lattice spacing. The quantity $\gamma$ encodes a horizontal stretch. It can be viewed as a decay constant -- the higher the value of $\gamma$, the more tightly bound is the wavefunction. 
As shown in Fig.~\ref{fig.2d_sing_bss} (bottom centre), $\gamma$ increases monotonically with $M$. Finally, Fig.~\ref{fig.2d_sing_bss} (bottom right) shows the binding energy vs. $M$. This encodes the energy difference between the bound state and the lowest delocalized state ($E_{min.delocalized,2D}=-4t$). As $M$ increases, the binding energy increases. This supports the contention that the ground state becomes more tightly bound.

We summarize these findings as follows. A bound state is formed for any singularity of co-dimension $2$. The higher the degree of the singularity, the more tightly bound the state.

\subsection{Comparison with bound states induced by a potential}

For comparison, we consider a smooth two-dimensional surface with a local attractive potential. The Hamiltonian for this system is similar to Eq.~\ref{eq.Hpot}. We have a single $L\times L$ sheet with an attractive potential of strength $g$ at the central site. Analytic solutions cannot be found, but a single bound state is seen in the numerics for any attractive potential. If the system size is large enough, the wavefunction decays to zero at the boundaries. As a result, the ground state is not sensitive to boundary conditions.
We describe this state below and compare with the continuum problem of a local attractive potential in two dimensions\cite{Nieto2002}. 
\begin{figure*}
\includegraphics[width=2.3in]{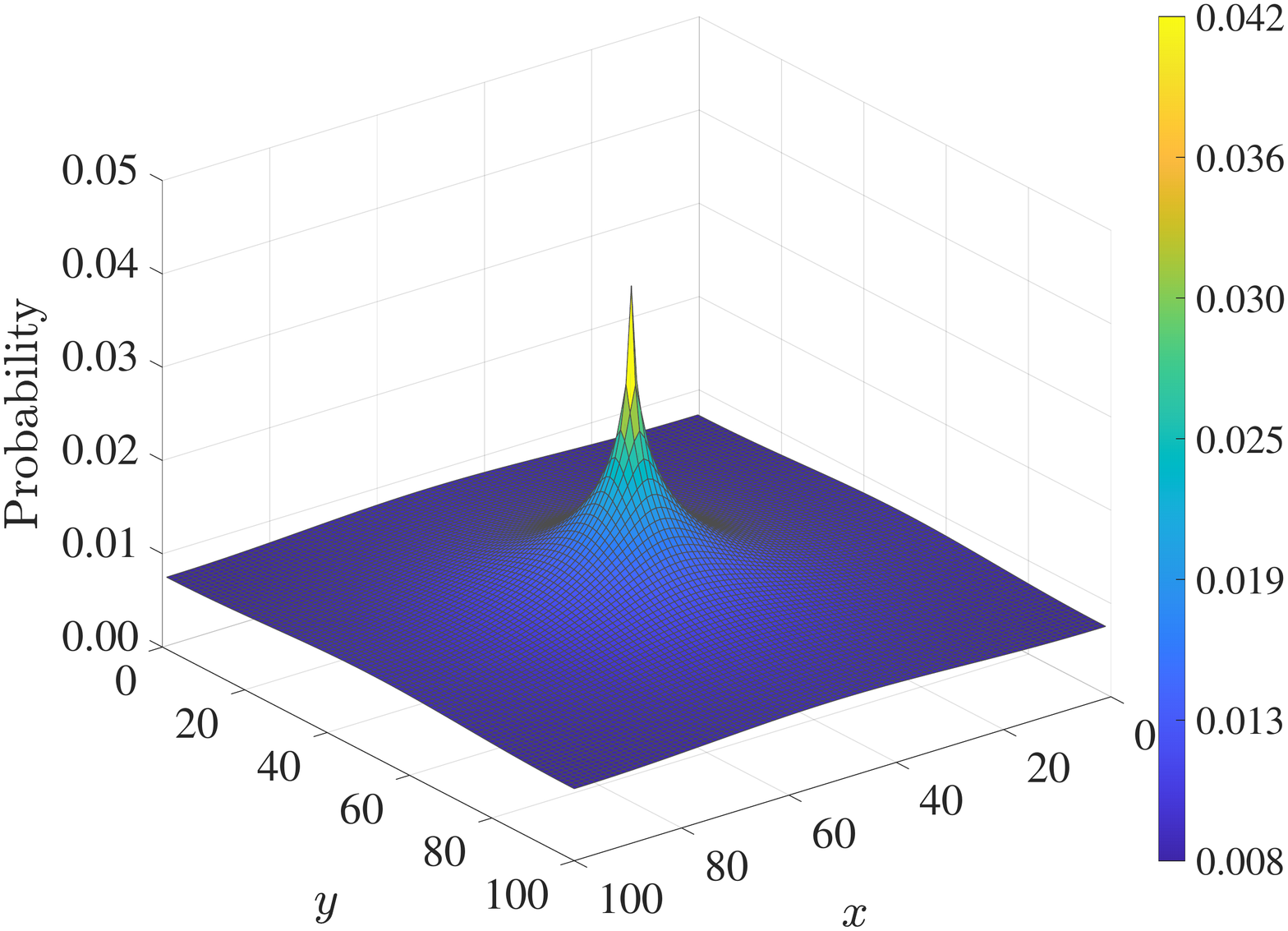}
\includegraphics[width=2.3in]{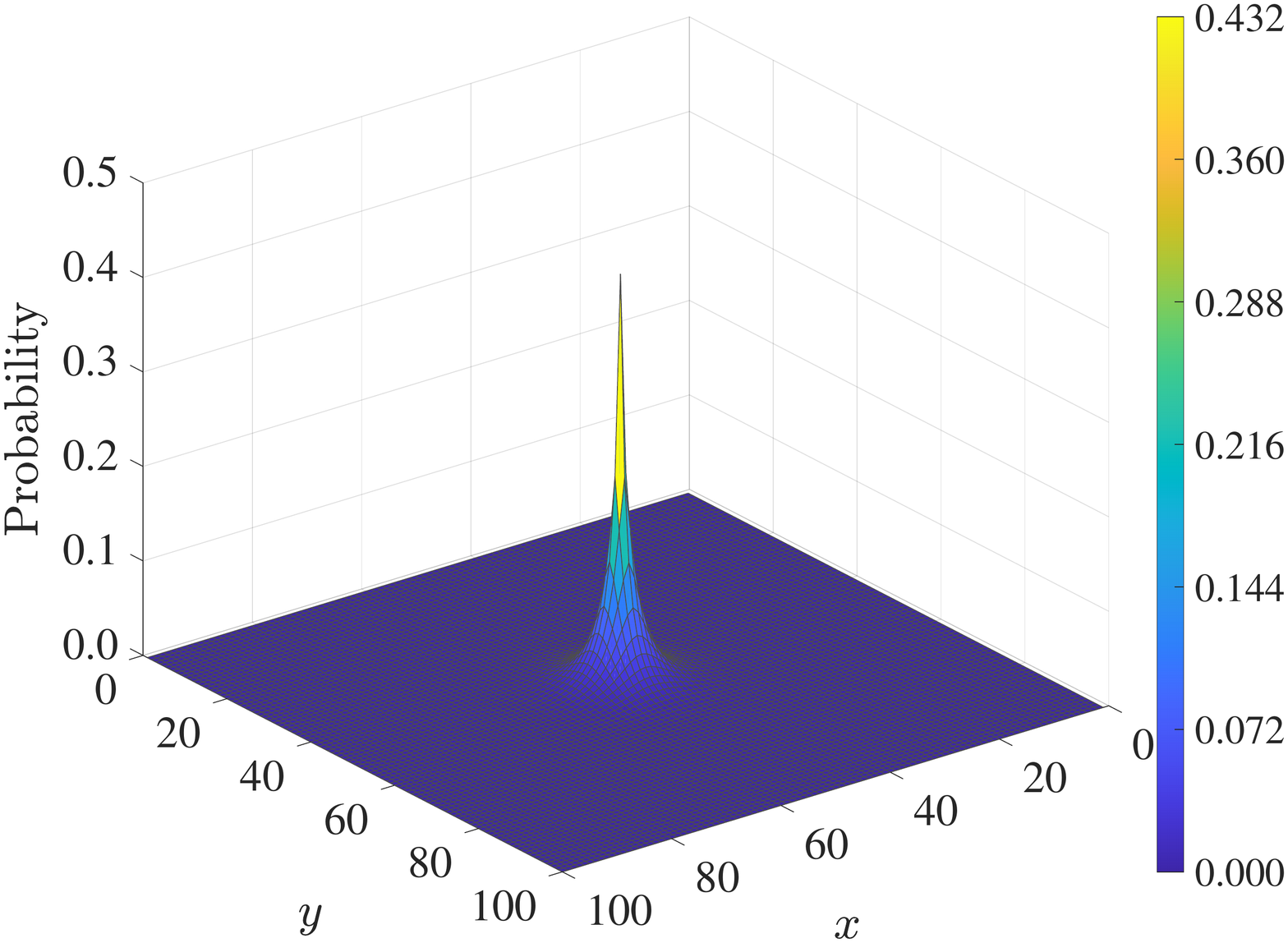}
\includegraphics[width=2.3in]{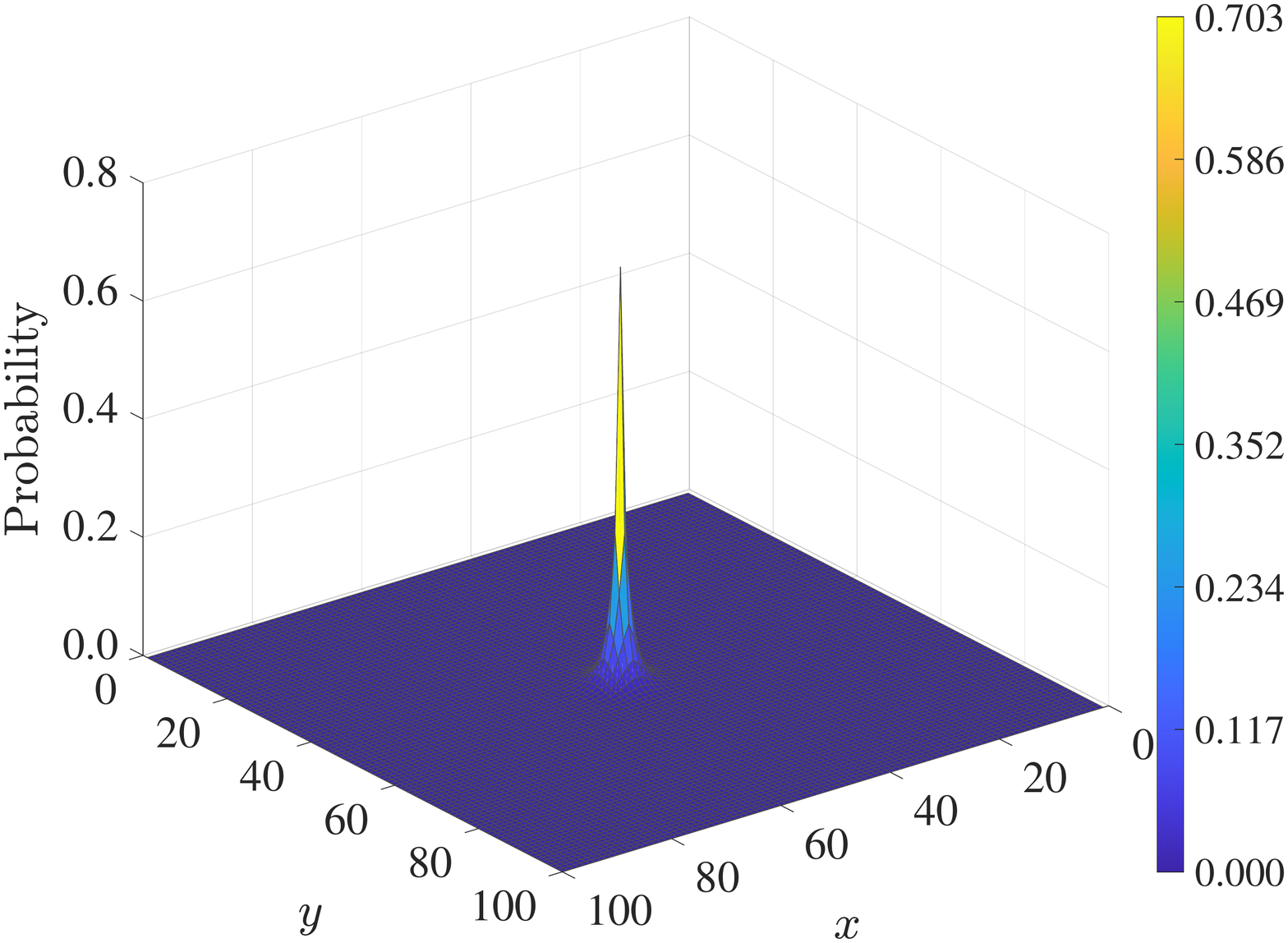}\\
\includegraphics[width=2.3in]{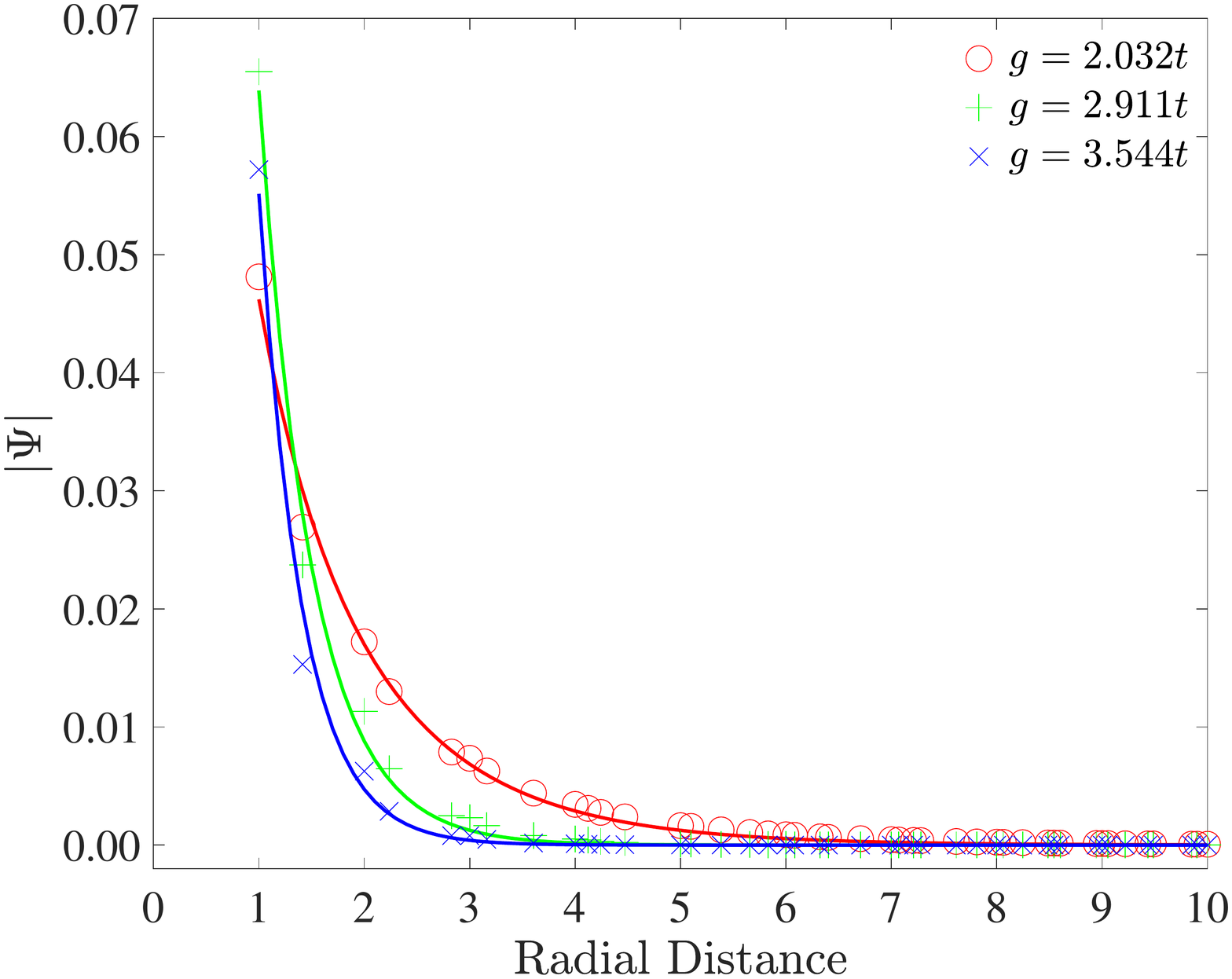}
\includegraphics[width=2.3in]{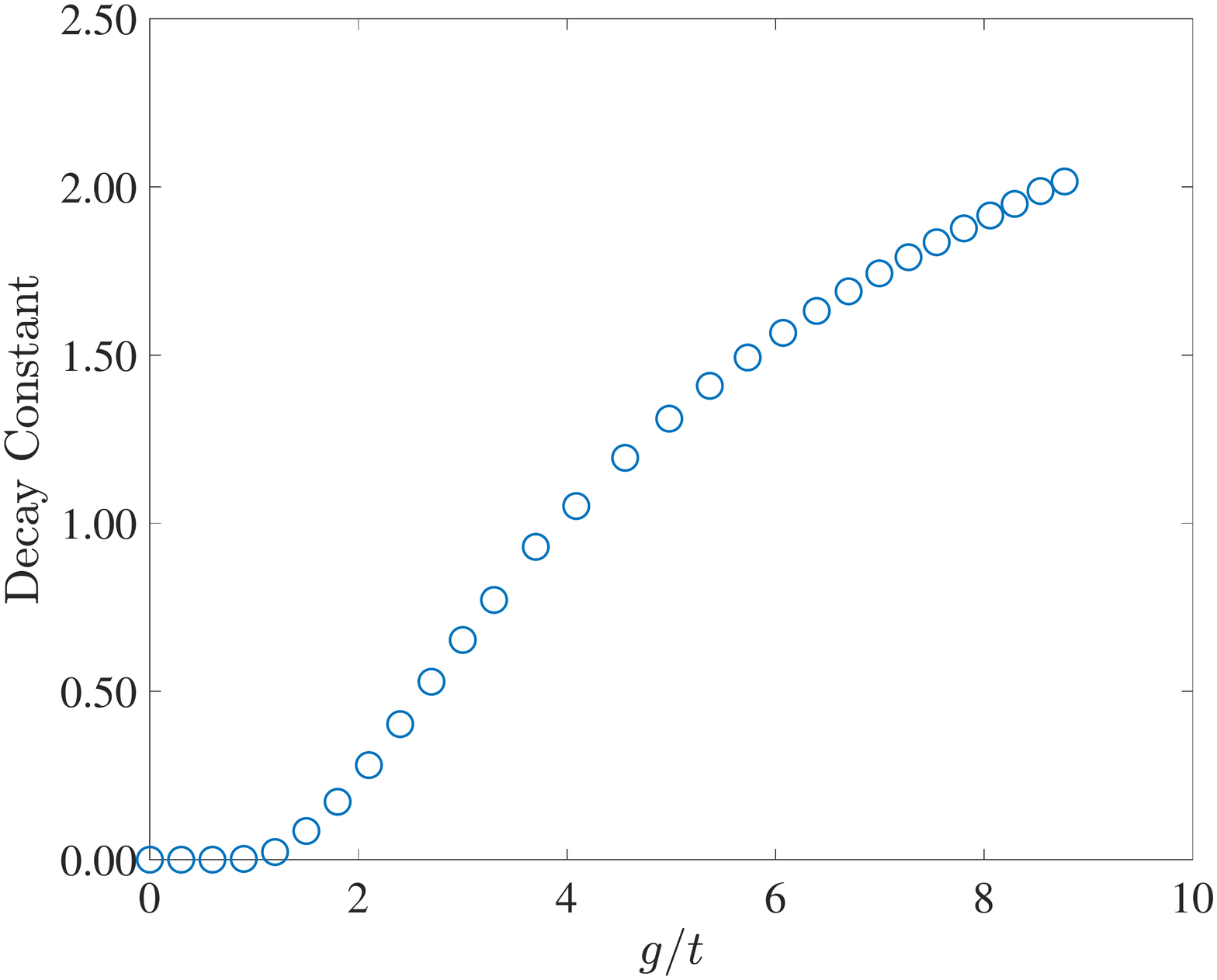}
\includegraphics[width=2.3in]{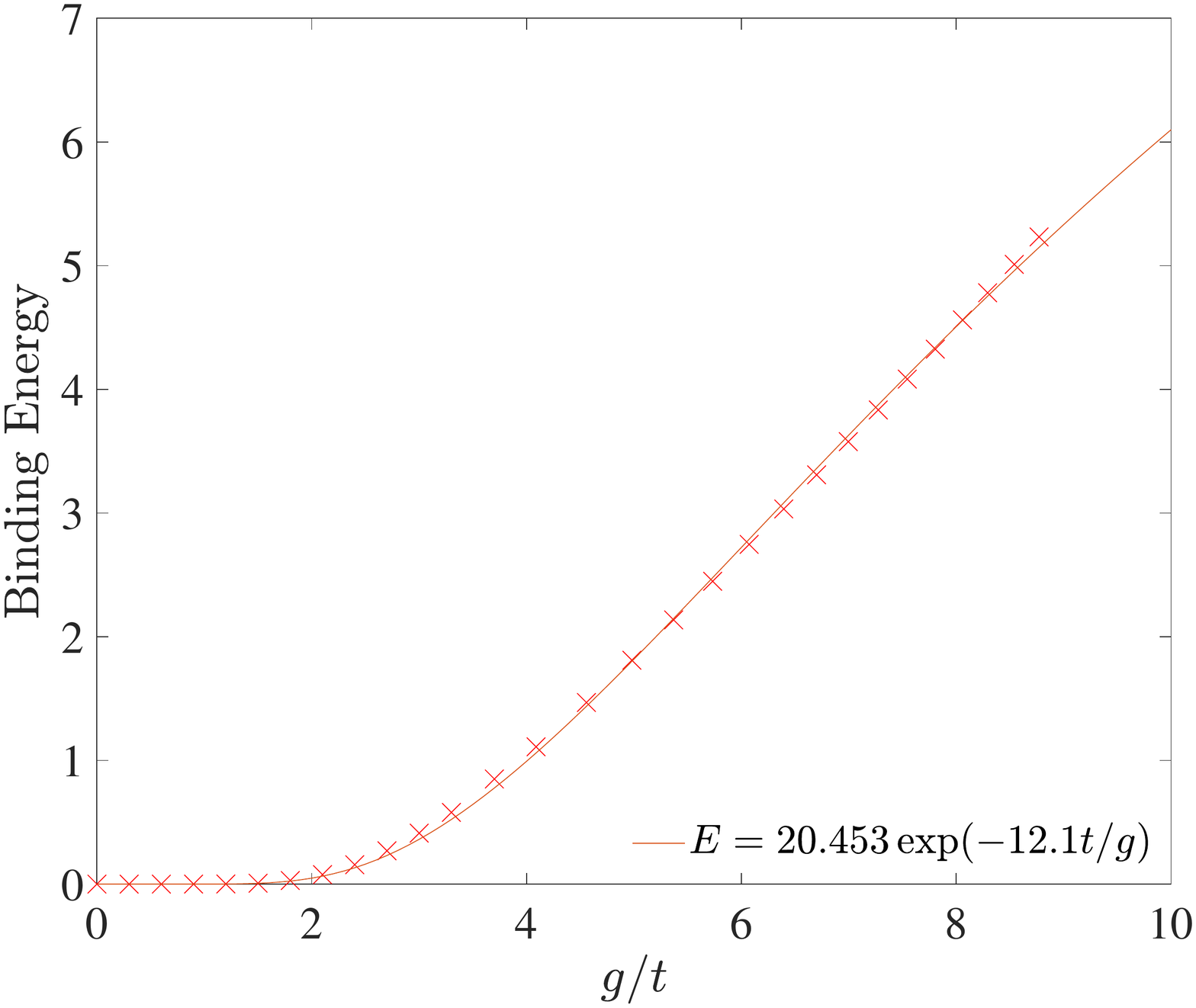}
\caption{The ground state wavefunction on a 2D sheet with an on-site attractive potential. The panels on top correspond to $g = t$ (top left), $g=2.032~t$ (top centre) and $g=2.911~t$ (top right). They show the spatial variation of the probability (amplitude squared). The 2D sheet is taken to be a 100 $\times$ 100 grid with periodic boundaries. The wavefunction profile is shown at bottom left, with the amplitude plotted against the radial coordinate. The data are fit to modified Bessel functions of order 0. 
At bottom centre, we plot the decay constant vs. $g$. At bottom right, we plot the binding energy (in units of $t$) vs. the strength of the potential. The best-fit curve to the form $E_{binding} = A e^{-B/g}$ is shown.} 
\label{fig.2d_pot_bss}
\end{figure*}

Fig.~\ref{fig.2d_pot_bss} (top) shows the bound state wavefunction for $g/t = 1,~2.032$ and $2.911$. The latter two values are chosen as they are equivalent to singularities with $M=2$ and $3$ respectively, as we discuss below. We emphasize that any value of $g$ produces a similar ground state. Fig.~\ref{fig.2d_pot_bss} (bottom left) shows the same wavefunctions, fitted to the form given in Eq.~\ref{eq.modBesform}. We find good agreement with the Bessel function form, especially at large distances. 
As shown in Figs.~\ref{fig.2d_pot_bss} (bottom centre, bottom right), the decay constant ($\gamma$) and the binding energy increase monotonically with $g$. The larger the potential, the tighter is the bound state. For small $g$, the binding energy is exponentially weak. As shown in the figure, the dependence on $g$ is well fit by the function $E_{binding} = A e^{-B/g}$. This form is known from the continuum problem of a bound state induced by a local potential (say, of the delta-function form). For example, it is invoked in the discussion of the Cooper instability\cite{Cooper1956,Esebbag1992}, where electron-pairs that are constrained to live on a two-dimensional space experience a weak attraction.

\begin{figure}
\includegraphics[width=3.1in]{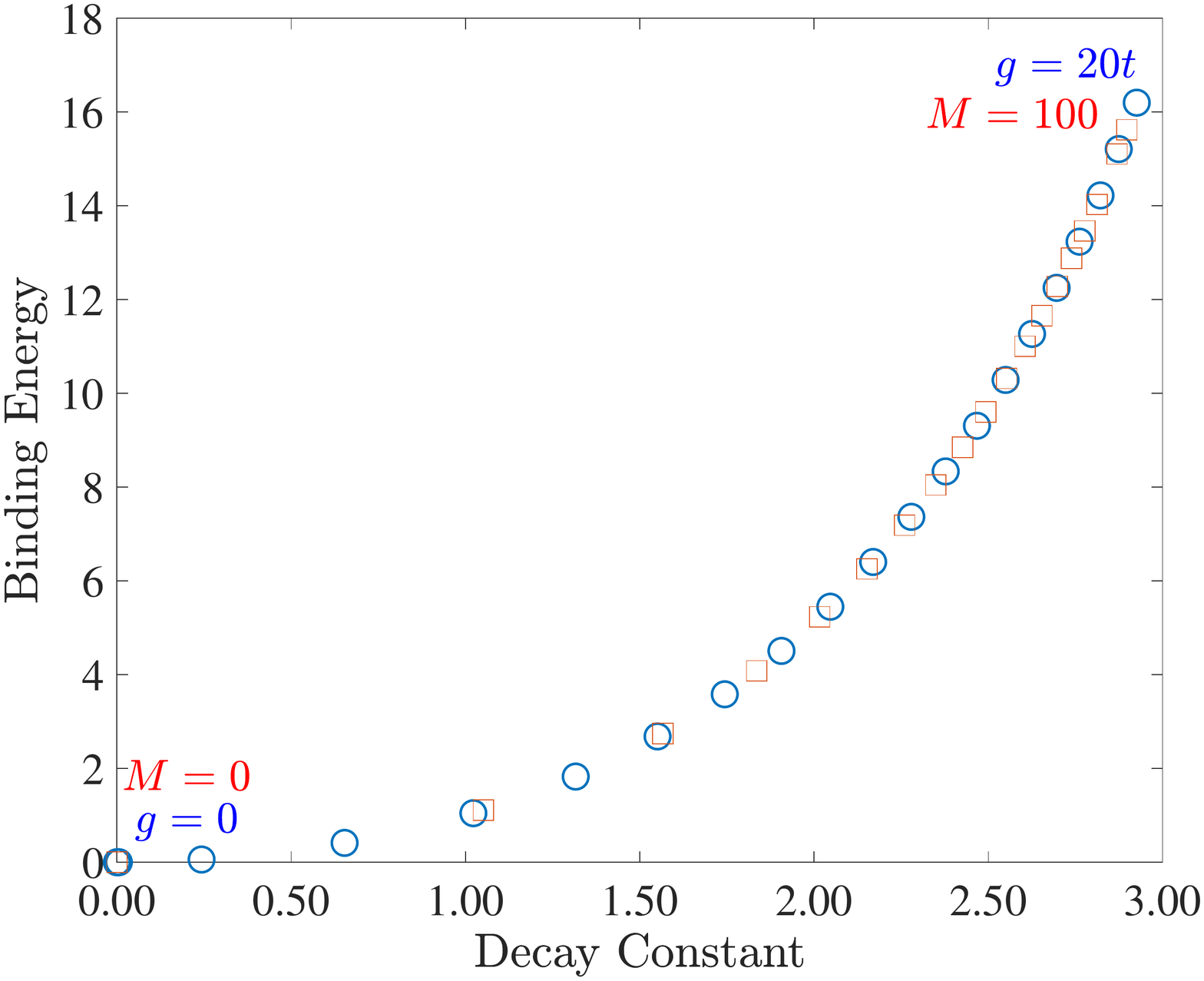}\\
\includegraphics[width=3.1in]{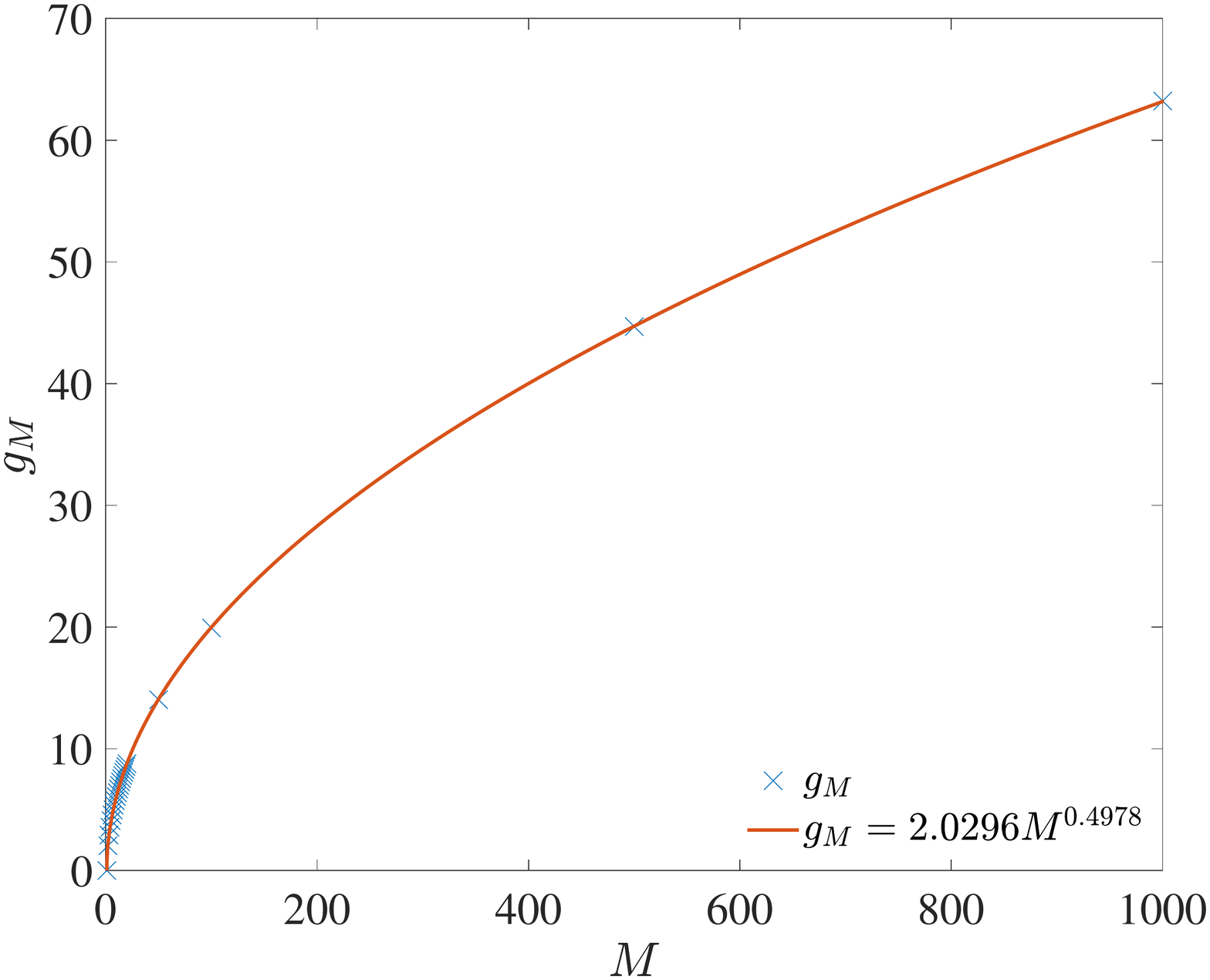}
\caption{Top: Binding energy vs. decay constant in two dimensions. Blue circles represent bound states induced by a local potential on a smooth two-dimensional sheet. From left to right, the points correspond to increasing potential strength, with $g$ increasing from $0$ to $20~ t$ in steps of $t$. Red squares represent bound states induced by singularities of co-dimension two. From left to right, the degree of the singularity ($M$) increases in steps of $5$ from $0$ to $100$. The two data sets collapse onto the same curve. 
Bottom: $g_{M}$ vs. $M$, where $g_{M}$ is the potential that is equivalent to a singularity of degree $M$. The data fit well to a function of the form $g_M \sim \sqrt{M}$.  } 
\label{fig.be_gamma}
\end{figure}

We now compare bound states induced by singularities with those induced by potentials. We treat $M$ and $g$ as tunable handles in the two cases. In both, we obtain localized ground states that fit well to the same functional form. In gross terms, the solutions are described by two parameters: the decay constant and the binding energy. These are both monotonically increasing functions of the tuning handle in each case. Remarkably, they are not independent. The decay constant immediately determines the binding energy and vice versa. This is shown in Fig.~\ref{fig.be_gamma} which plots the binding energy vs. decay constant for both singularity-induced and potential-induced bound states. The data points collapse onto the same curve. This leads us to conclude that potentials and singularities lead to the same bound states. For a singularity of degree $M$, we can find an equivalent potential $g_M$ that generates a bound state with the same decay constant and binding energy. Fig.~\ref{fig.2d_pot_bss} (bottom right) shows the variation in $g_M$ with $M$. As $M$ increases, the equivalent potential grows in strength. For large $M$, we find $g_M \sim \sqrt{M}$. 

We have verified that the equivalence goes beyond the decay constant and binding energy. It holds even for the precise form of the wavefunction, up to a change in the normalization to account for multiple sheets.

\section{$D=3$: Intersecting three-dimensional spaces}

We proceed to singularities of co-dimension 3, with an example shown in Fig.~\ref{fig.spaces} (right). As with co-dimension-2, analytic solutions cannot be found. We present numerical solutions and fit them to functional forms that are inspired by the continuum problem.

\subsection{Numerical solution to the tight-binding problem}

We consider $M$ three-dimensional spaces that share a common point. We discretize each space using an $L \times L \times L$ cubic grid. The central point is taken to be common to all spaces. While a generic point has 6 neighbours, the centre has $6M$ neighbours. This configuration defines a graph with $\{ML^3 - (M-1)\}$ sites. We solve the tight binding Hamiltonian of Eq.~\ref{eq.Hgen} on this graph numerically. We examine the energy and wavefunction of the lowest eigenstate. If a state is bound and $L$ is large enough, the wavefunction will decay before reaching the boundaries. The wavefunction will then be indifferent to open or periodic boundary conditions.

\begin{figure*}
\includegraphics[width=2.3in]{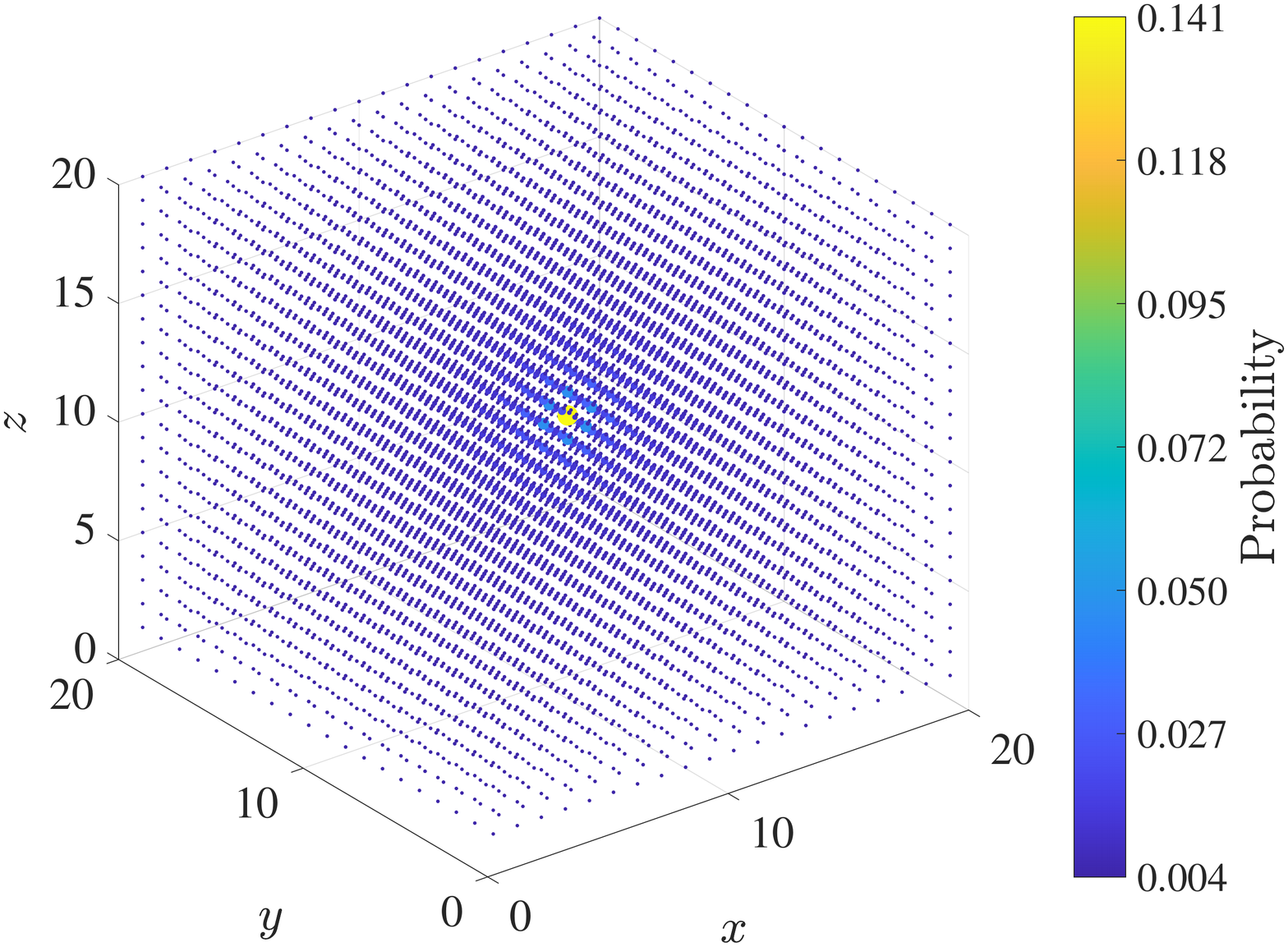}
\includegraphics[width=2.3in]{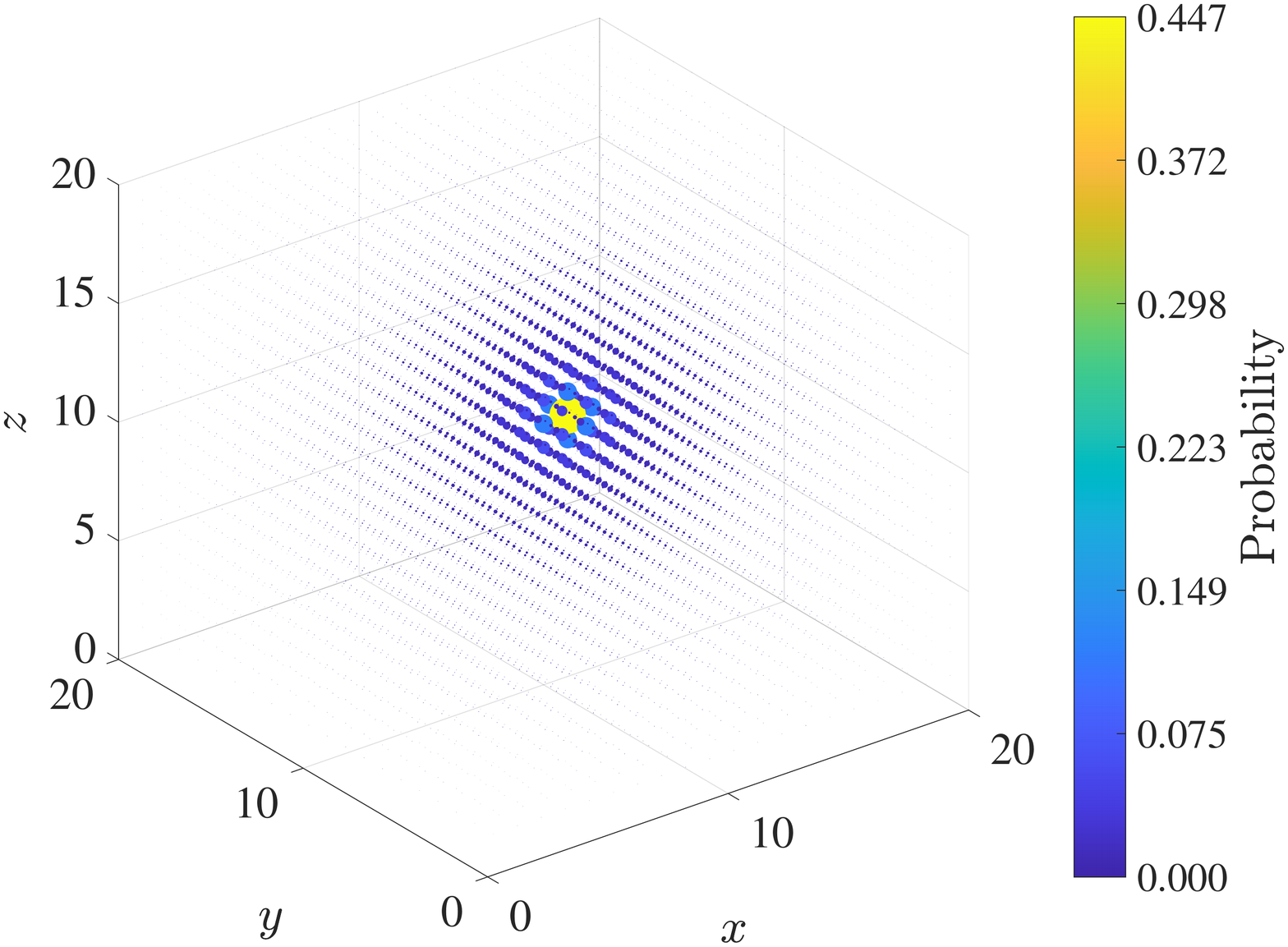}
\includegraphics[width=2.3in]{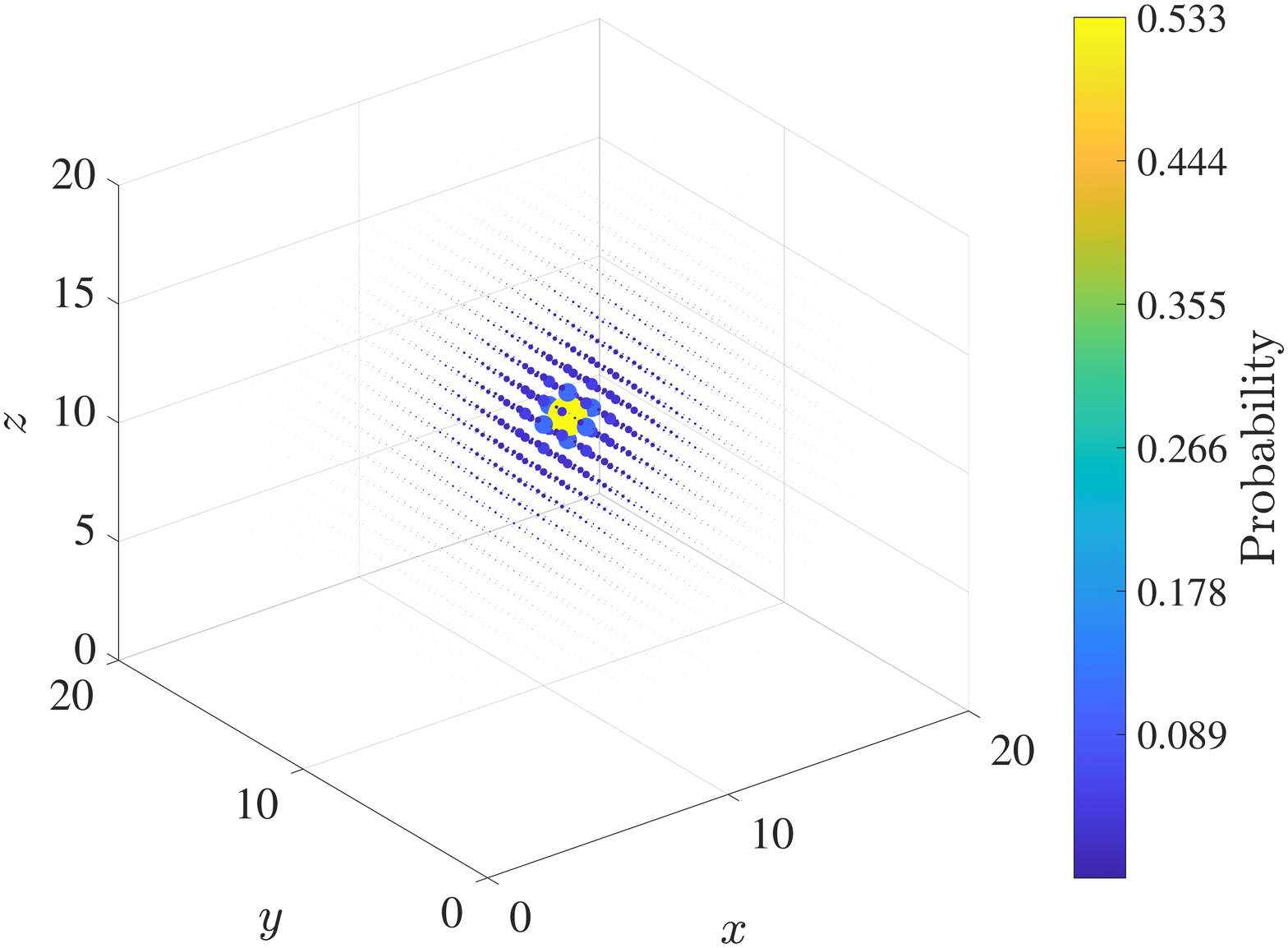}\\
\includegraphics[width=2.3in]{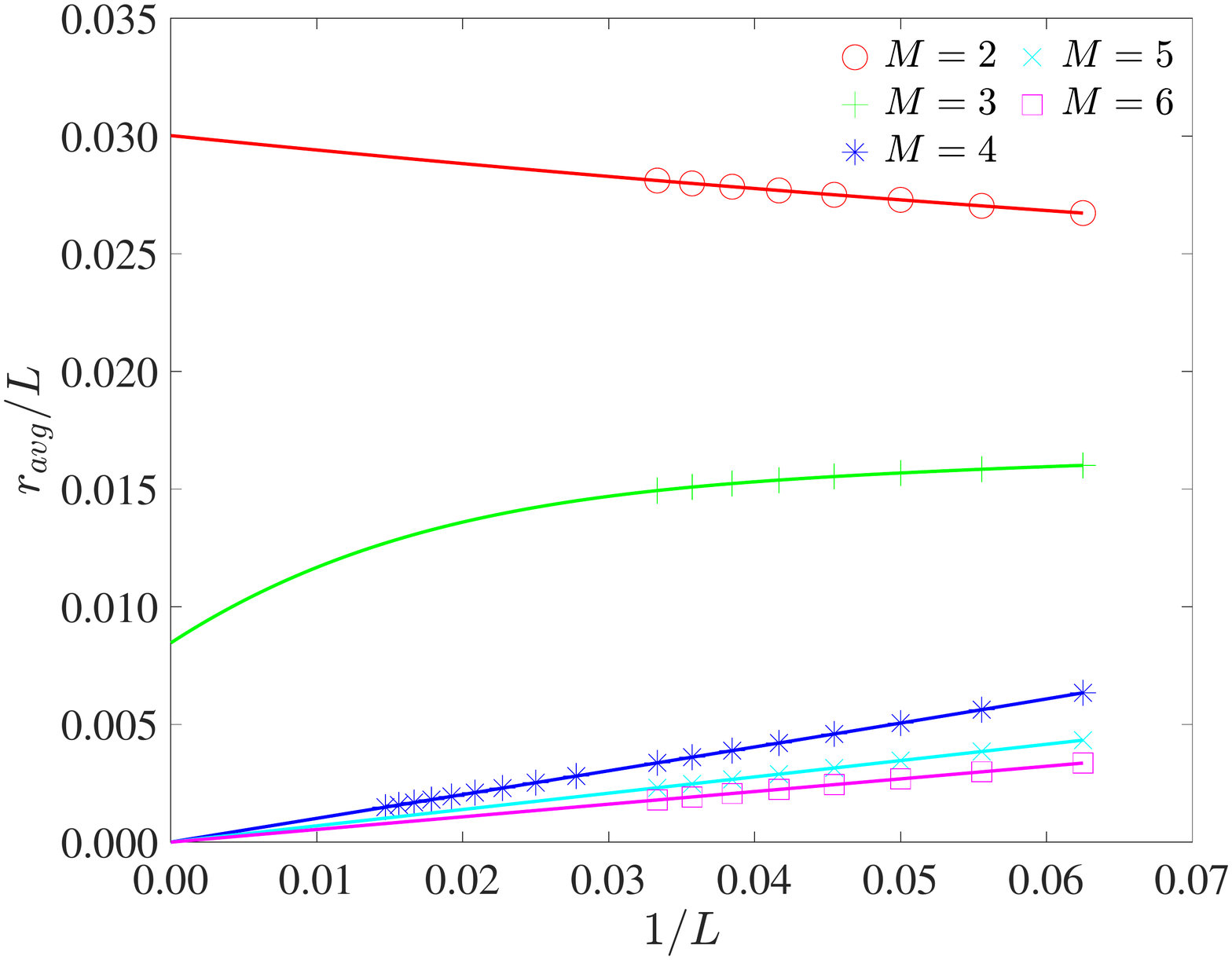}
\includegraphics[width=2.3in]{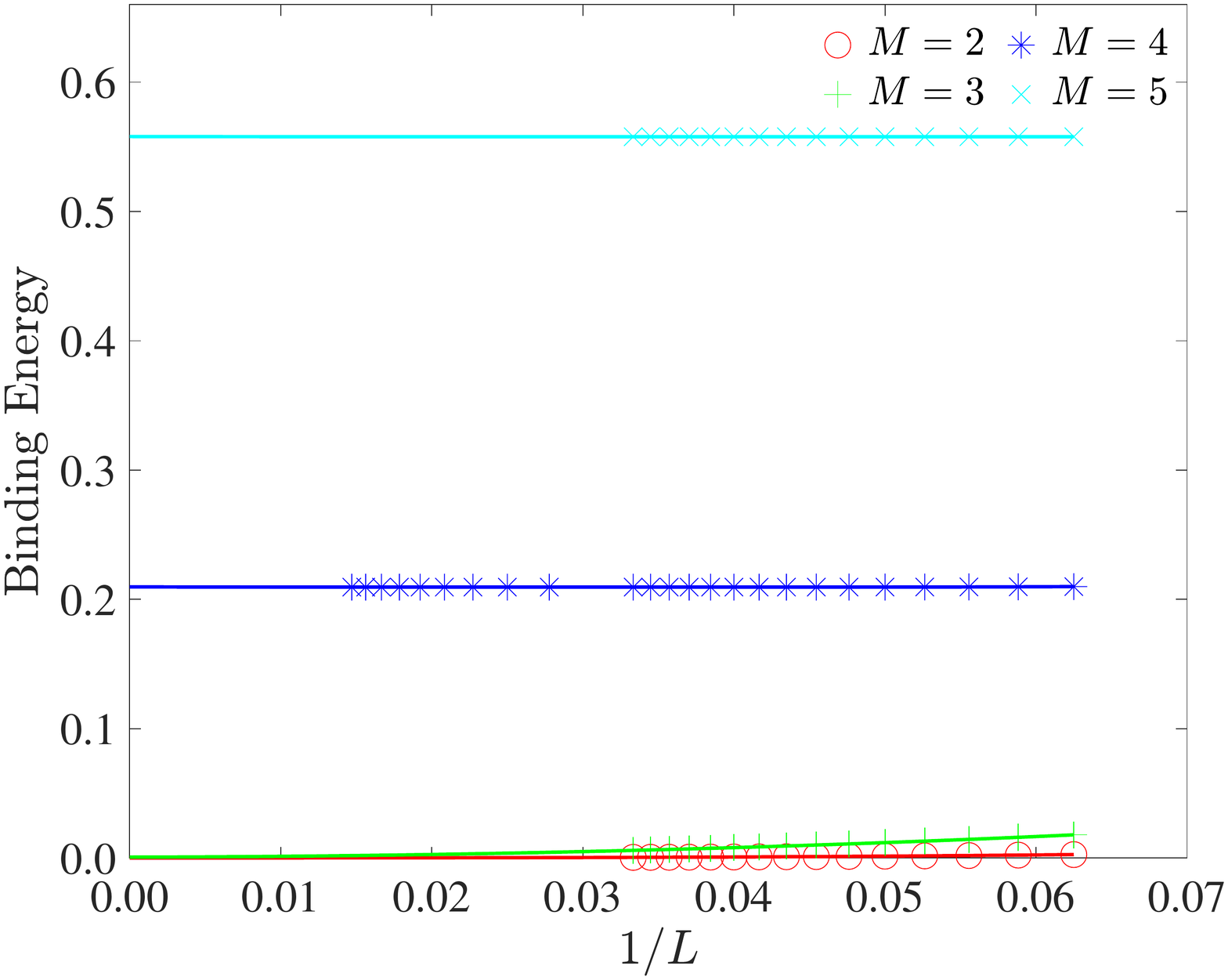}
\includegraphics[width=2.3in]{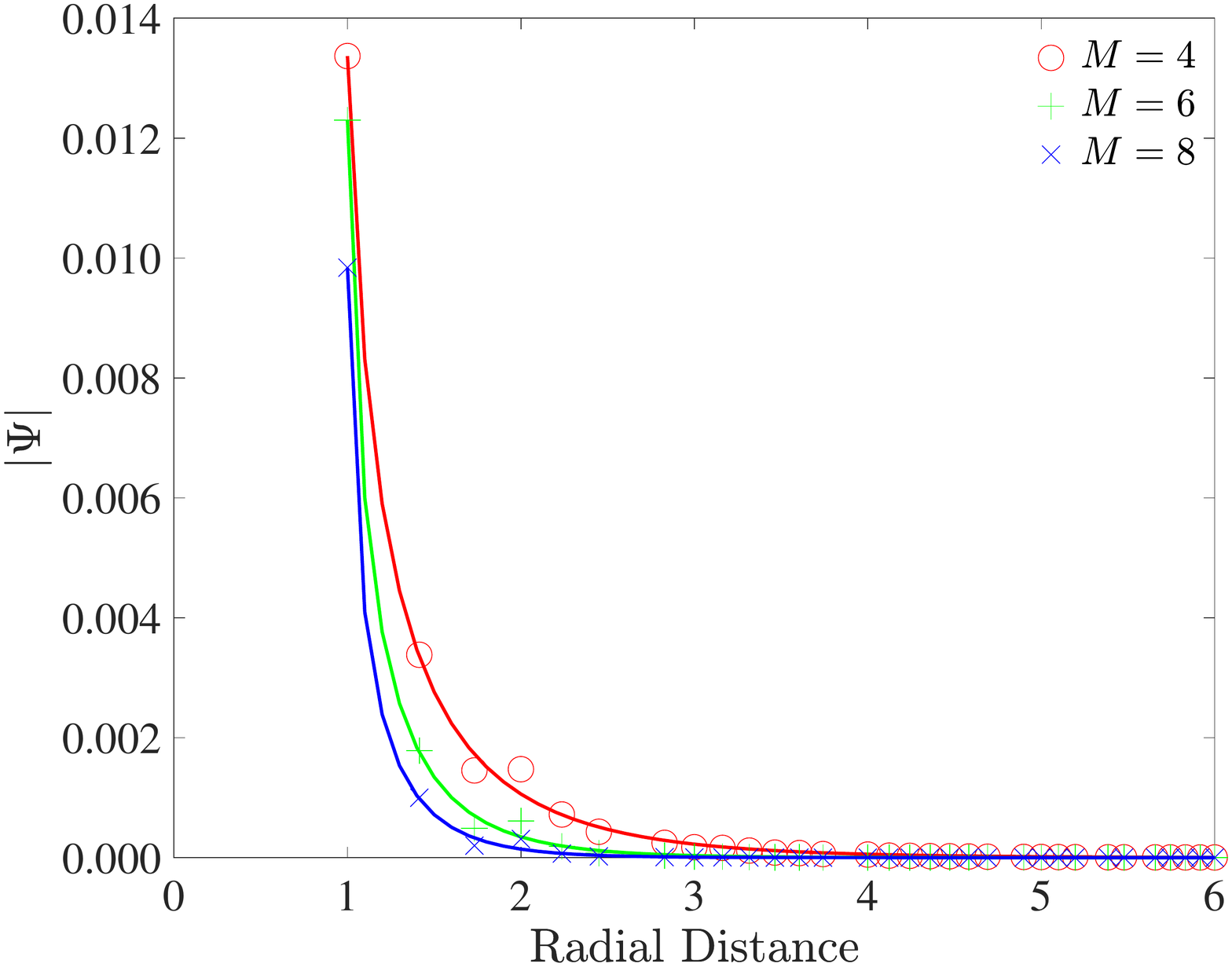}
\caption{The ground state of a particle on a space of $M$ cubes intersecting at a point, where $M=3$ (top left), $M=4$ (top centre) and $M=5$ (top right). Each cube is taken to be a 20 $\times$ 20 $\times$ 20 grid with periodic boundaries. The centre point is shared by all cubes, representing a point of intersection. In each plot, the size and colour of markers both represent the probability of finding the particle (amplitude squared) at a certain site. We only show sites on a single cube as the wavefunction is the same across cubes. At bottom left, we plot $r_{avg}/L$ vs. $1/L$ (see text). The data are fit to the form $y = a + b/L + c/L^2$ in order to extrapolate to $L \rightarrow \infty$. At bottom centre, we plot the binding energy vs. $1/L$ with fits to the form $y = a + b/L + c/L^2$. Finally, at bottom right, we plot wavefunction amplitude against the radial coordinate for $M=4,6,8$. The data are fit to a modified Bessel form (see text). }
\label{fig.3d_bss}
\end{figure*}

The co-dimension-3 solutions present a remarkable difference when compared with co-dimensions 1 and 2. A bound state forms only when the degree of the singularity exceeds a threshold value. For example, as shown in Fig.~\ref{fig.3d_bss} (left), we find no bound state for $M=3$ (three cubes intersecting a point). However, there is a clear bound state when $M=5$ (five cubes intersecting at a point), as shown in Fig.~\ref{fig.3d_bss} (right). This can be seen in various ways as we describe below. We first note that any such analysis requires a systematic approach to the thermodynamic limit by increasing $L$. A true bound state will remain bound with a constant `width' as $L$ increases. In contrast, a delocalized state will expand with increasing system size.  

We introduce a quantitative measure for localization,
\bea
r_{avg} = \sum_{j} \vert \vec{r}_j - \vec{r}_0 \vert \times \vert \psi_{j,0} \vert^2,
\eea
where $j$ runs over all sites in the tight-binding setup. The distance between the origin and site $j$ is denoted as $\vert \vec{r}_j - \vec{r}_0 \vert$. Note that distances are calculated as on the usual cubic lattice: a point with coordinates $(x,y,z)$ is at a distance of $\sqrt{x^2 + y^2 + z^2}$ from the origin. The probability amplitude of the \textit{ground state} at this site is denoted as $\vert \psi_{j,0} \vert$. Assuming that the particle resides in the ground state, $r_{avg}$ denotes its average separation from the singularity. 
 If the ground state represents a true bound state, $r_{avg}/L$ will extrapolate to zero as $L\rightarrow \infty$. Instead, if the ground state were delocalized, $r_{avg}/L$ will extrapolate to a non-zero value. Fig.~\ref{fig.3d_bss} (bottom left) compares $r_{avg}/L$ vs. $1/L$ for various $M$ values. We see a qualitative shift between $M=3$ and $M=4$, with $M \geq 4$ showing bound state formation. 

We next examine the binding energy. It is defined as the energy separation between the lowest state and the bottom of the delocalized continuum ($E_{min.delocalized,3D}=-6t$). If the ground state were truly bound, $E_{binding}$ will approach a non-zero value as $L\rightarrow \infty$. In a delocalized state, $E_{binding}$ will vanish for large $L$. Fig.~\ref{fig.3d_bss} (bottom centre) shows $E_{binding}$ vs. $1/L$ for various $M$ values. Once again, we find behaviour that is consistent with bound state formation only when $M \geq 4$.

To characterize the wavefunction in a systematic manner, we fit it to the form 
\bea
h(r)= c~r^{-1/2} k_{1/2}(\gamma r + b).
\label{eq.modBessform3d}
\eea
Here, $k_{1/2}$ represents a modified Bessel function of the second kind of order $1/2$. This form is known from the continuum problem of a 3D attractive square well. When a bound state is produced, its wavefunction follows this form in the external region (outside the well)\cite{Nieto2002,Geltman2011}. We find a good fit to the modified Bessel form as long as $M \geq 4$, as shown in Fig.~\ref{fig.3d_bss} (bottom right). The wavefunction amplitude at each site depends solely on the radial distance from the singularity. The phase is ignored as it is uniform at all sites.

From the fit, we obtain the decay constant $\gamma$. For $M \geq 4$, the decay constant increases with increasing $M$, as does the binding energy, as we discuss below. The ground state becomes progressively more bound.

\subsection{Comparison with bound states induced by a potential}

\begin{figure*}
\includegraphics[width=2.3in]{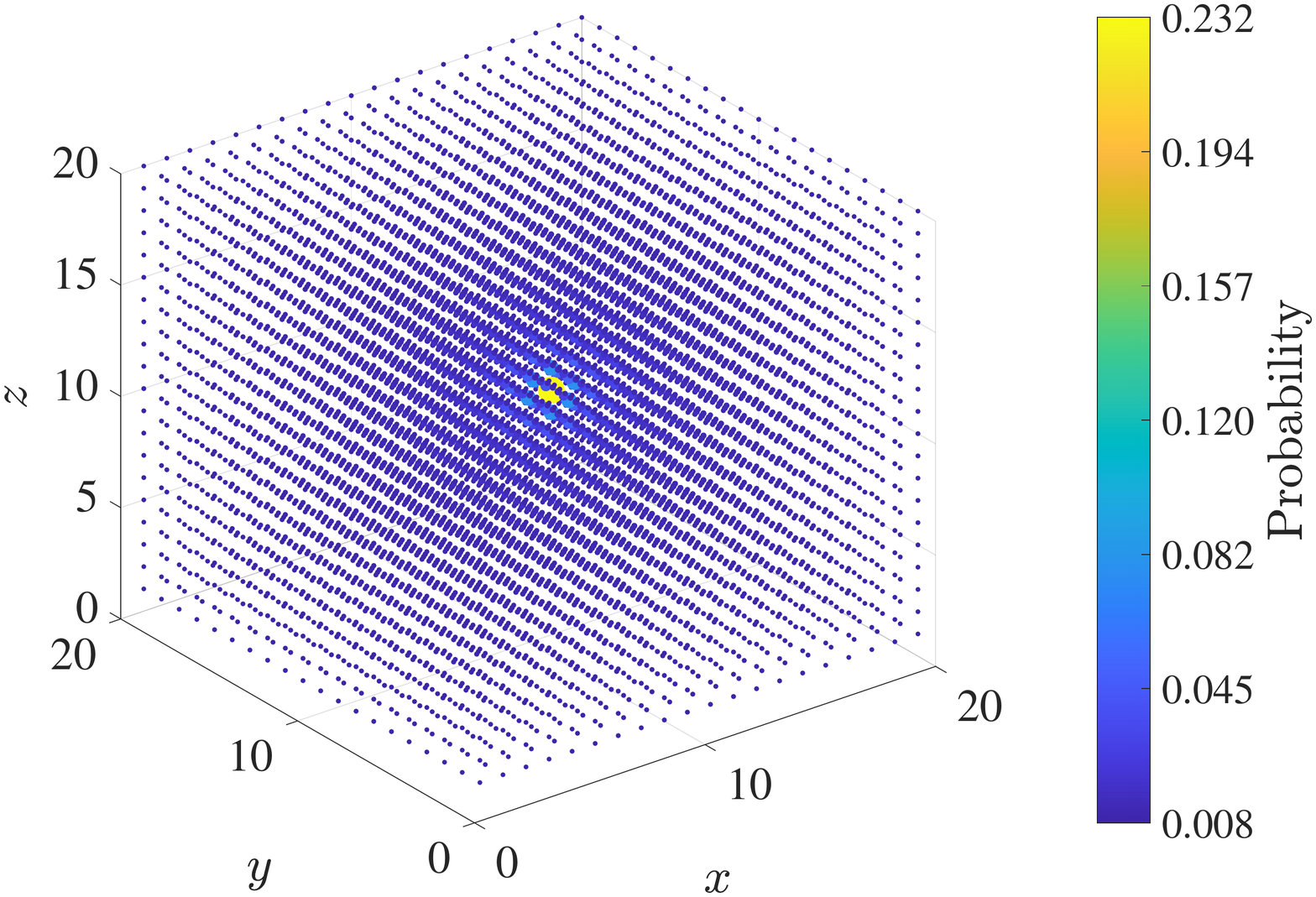}
\includegraphics[width=2.3in]{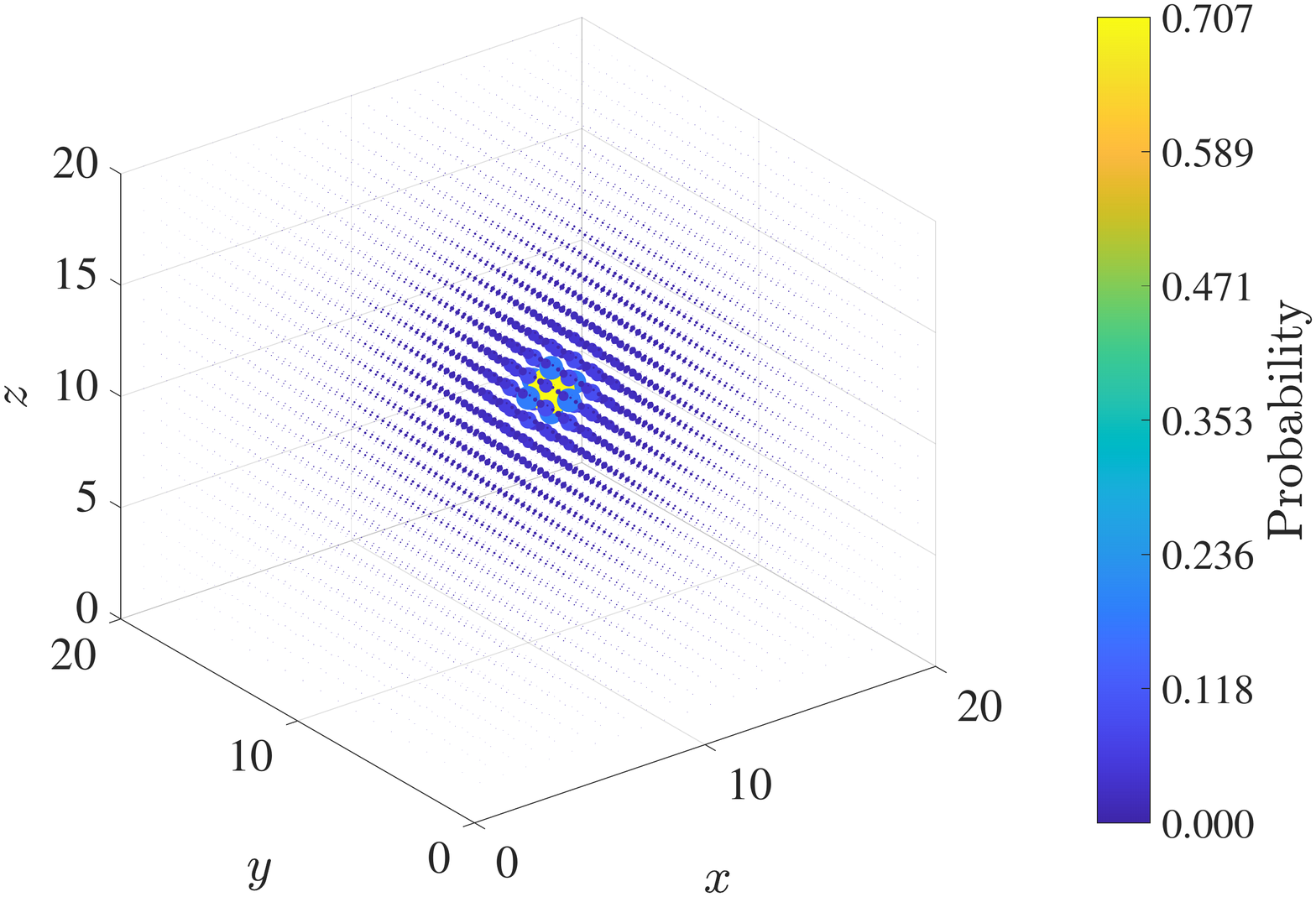}
\includegraphics[width=2.3in]{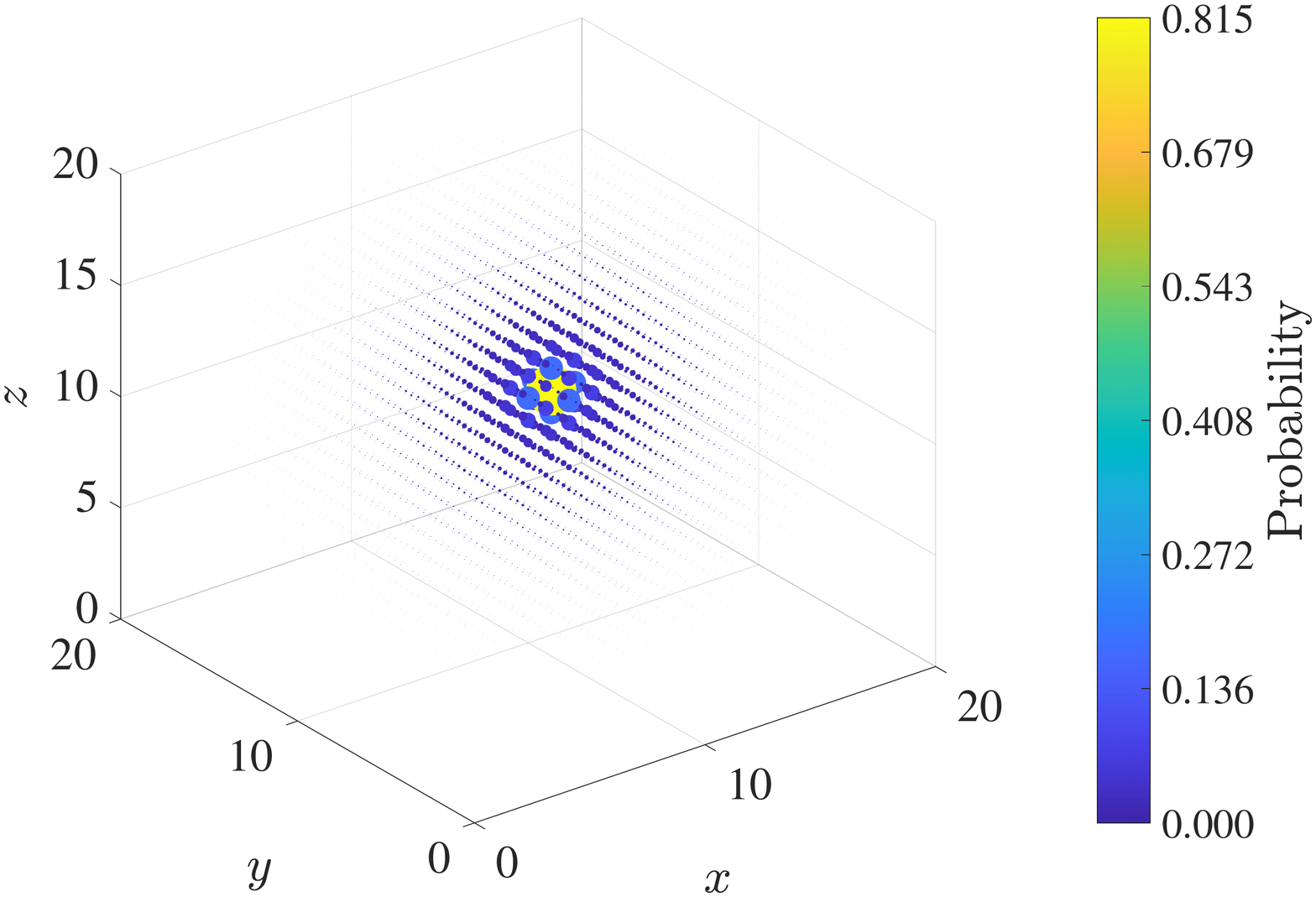}\\
\includegraphics[width=2.3in]{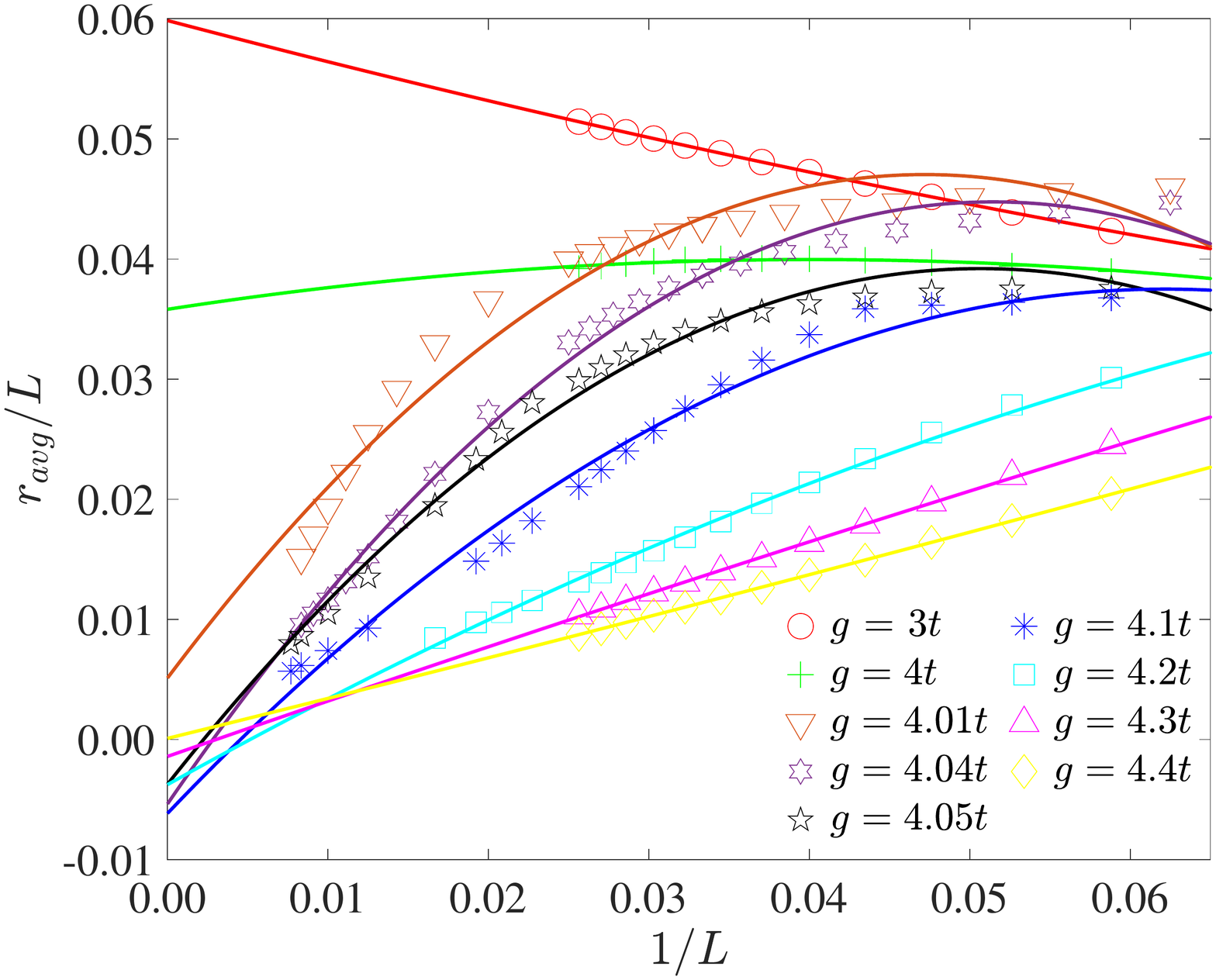}
\includegraphics[width=2.3in]{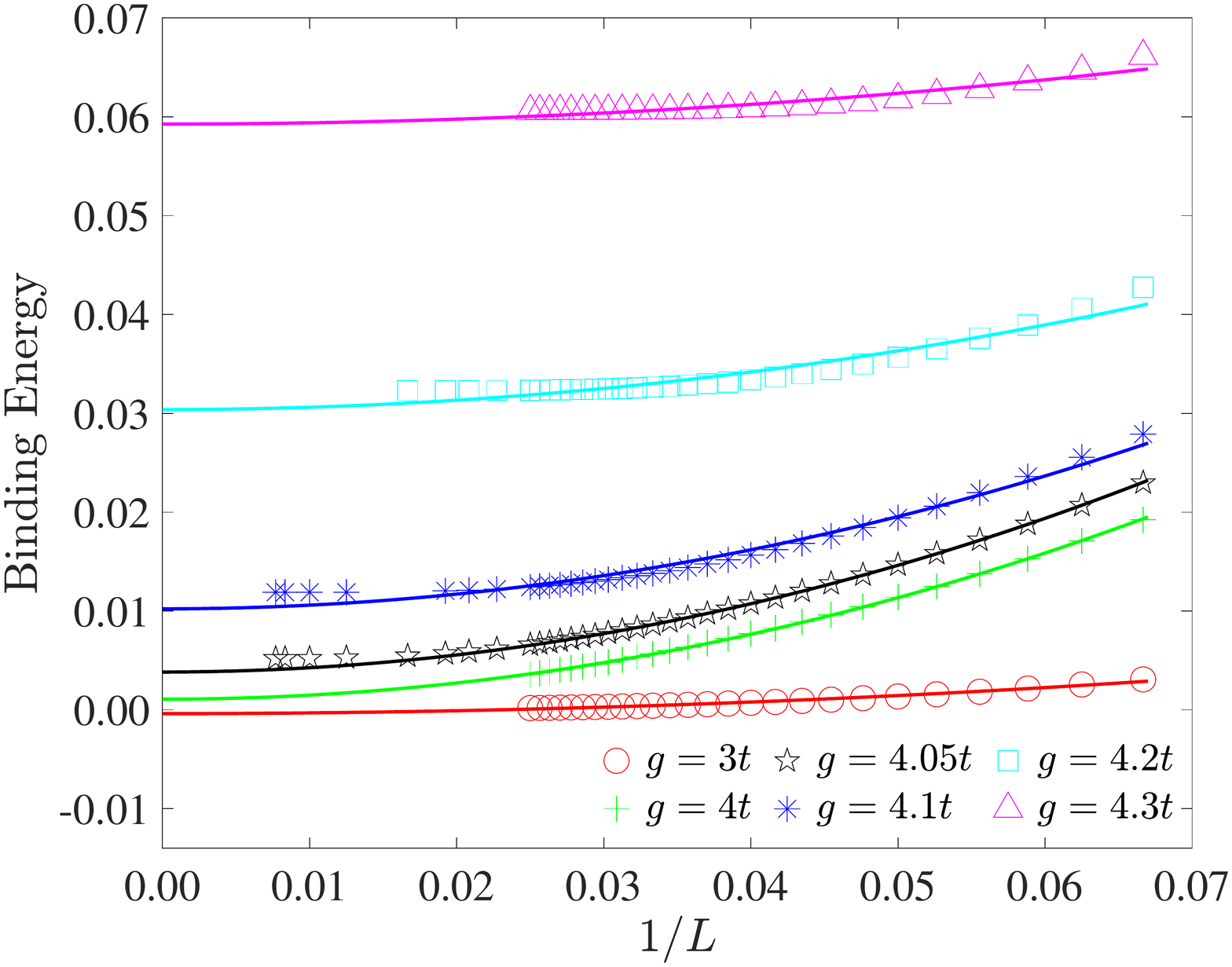}
\includegraphics[width=2.3in]{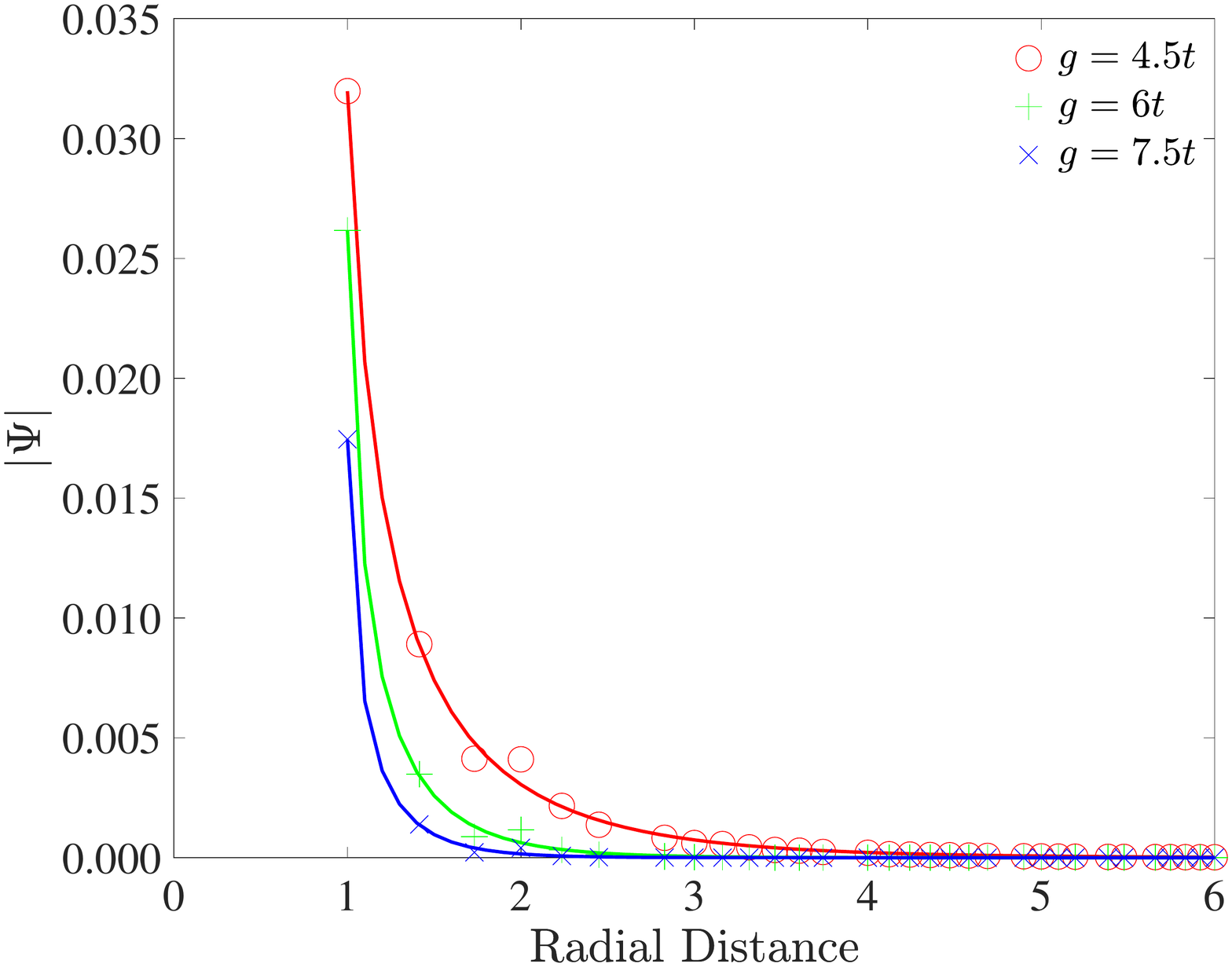}
\caption{The ground state on a 3D space with a local attractive potential. The panels on top correspond to $g=4~t$ (top left), $g=4.657 ~t$ (top centre) and $g=5.246~ t$ (top right). The space is taken to be a 20 $\times$ 20 $\times$ 20 cubic grid with periodic boundaries with the on-site potential at the centre. In each plot, the size and colour of markers both represent the probability of finding the particle (amplitude squared) at a certain site. At bottom left, we plot $r_{avg}/L$  vs. $1/L$ for various potential strengths. The data are fit to the form $y = a + b/L + c/L^2$. At bottom centre, we plot the binding energy (see text) vs. $1/L$, with fits to the form $y = a + b/L^2$. At bottom right, the wavefunction amplitude is plotted as a function of the radial distance from the potential. The data are fit to a modified Bessel function form (see text).} 
\label{fig.3d_pot}
\end{figure*}

As before, we compare our results with a smooth three-dimensional surface with a local attractive potential. We model this as a tight binding problem on an $L\times L\times L$ cubic grid. We place an attractive on-site potential of strength $g$ at the centre. 
For small values of $g$, the ground state is not localized. A bound state is formed only when $g$ exceeds a threshold value. The ground state is plotted in the top panels of Fig.~\ref{fig.3d_pot} for $g = 4~t$, $4.657~t$ and $5.246~t$. The first clearly shows a delocalized ground state, while the latter two are bound. In fact, we will argue below that the latter two values are equivalent to $M=4$ and $5$.

We approach the question of bound state formation in the same manner as with the singularity above. Figs.~\ref{fig.3d_pot} (bottom left) and (bottom centre) show $r_{avg}/L$ and the binding energy vs. $1/L$ for a few potential strengths. We see two clear regimes, $g\lesssim 4t$ and $g\gtrsim 4.05t$. In the former, $r_{avg}/L$ extrapolates to a non-zero value as $L\rightarrow\infty$. At the same time, the binding energy extrapolates to zero. In the latter, $r_{avg}/L$ extrapolates to zero, while the binding energy extrapolates to a non-zero value. This suggests a threshold value, $g_{critical}$, that separates bound and unbound behaviour. This result is the tight-binding analogue of a well-known result in quantum mechanics: in three dimensions, a critical potential strength is required for bound state formation.

The precise location of the critical point is difficult to pinpoint due to system size limitations. For example, near the transition, the fitting curves to $r_{avg}/L$ vs. $1/L$ cannot be reliably extrapolated to $1/L \rightarrow 0$ within accessible system sizes. The binding energy curves of Fig.~\ref{fig.3d_pot} (bottom centre) are somewhat clearer: $g_{critical}$ appears to lie between $g = 4.00~t$ and $4.05~t$.

 Fig.~\ref{fig.3d_pot} (bottom left) shows the evolution of $r_{avg}/L$ with $g$. We see a qualitative change between two regimes, one where $r_{avg}/L$ extrapolates to zero as $L \rightarrow \infty$ and the other where it tends to a non-zero value. The boundary between these regimes cannot be precisely discerned within accessible system sizes. In Fig.~\ref{fig.3d_pot} (bottom centre), we see the evolution of binding energy with system size, for various values of $g$. Based on the values extrapolated to $L\rightarrow \infty$, we conclude that the critical potential strength lies between $g = 4.00~t$ and $4.05~t$. Fig.~\ref{fig.3d_pot} (bottom left) shows the evolution of $r_{avg}/L$ with $g$.

When the potential exceeds $g_{critical}$, the resulting bound state fits well to the modified Bessel form of Eq.~\ref{eq.modBessform3d}. This is shown in Fig.~\ref{fig.3d_pot} (bottom right). The best-fit value of the decay constant, $\gamma$, increases monotonically with increasing $g$. Likewise, the binding energy increases with $g$. A bound state forms when $g$ exceeds $g_{critical}$, becoming progressively more bound as $g$ increases further.

\begin{figure}
\includegraphics[width=3.1in]{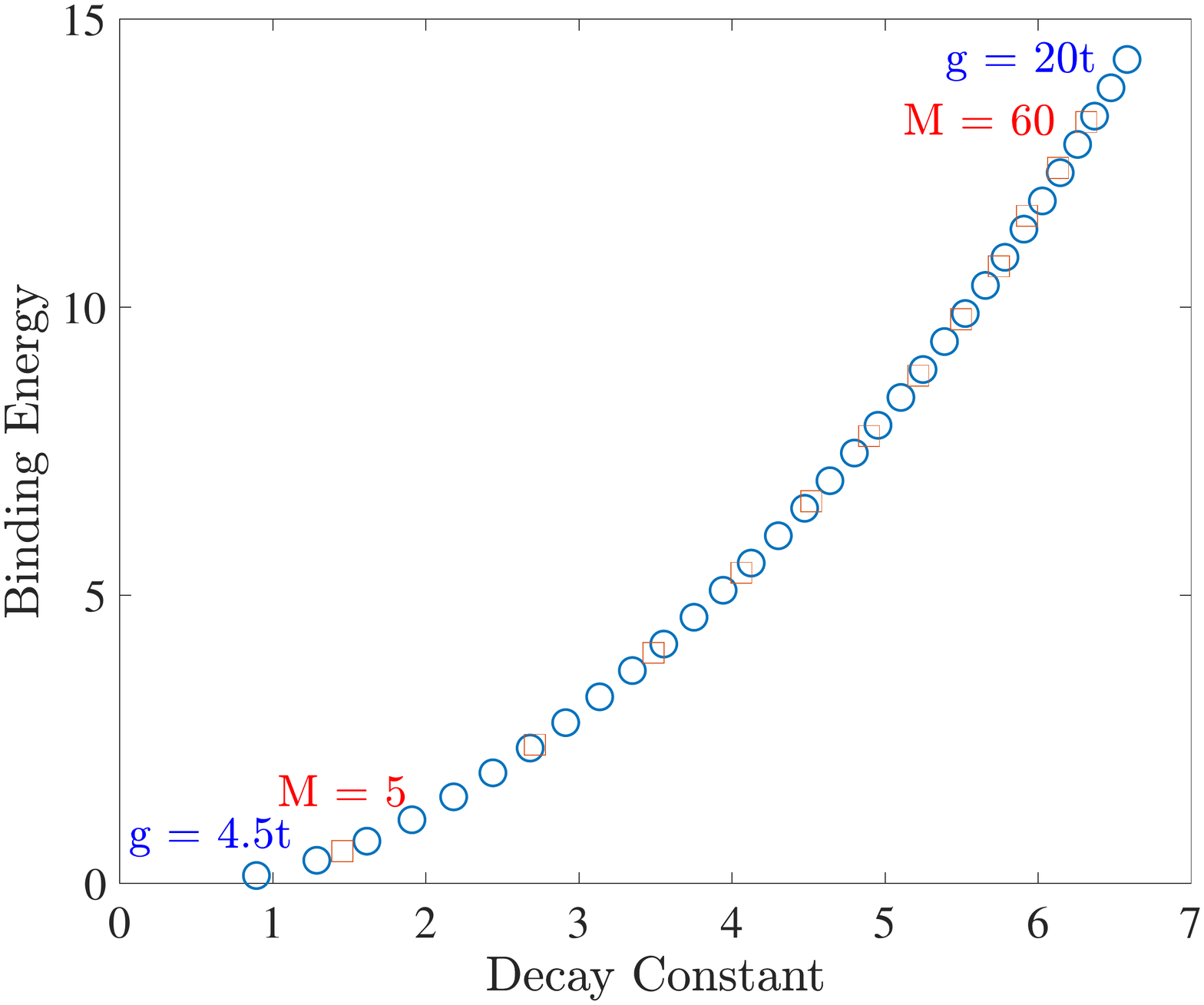}\\
\includegraphics[width=3.1in]{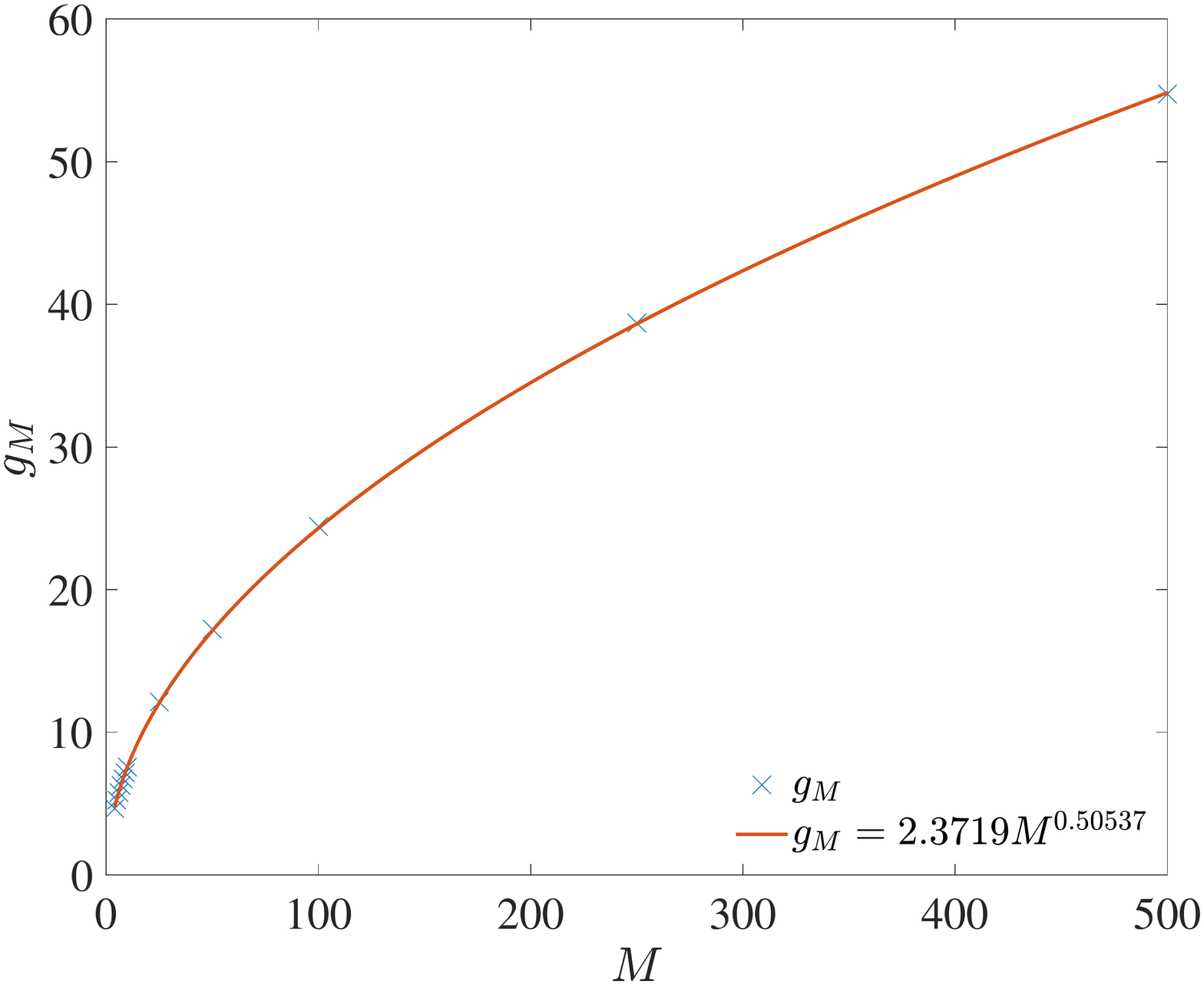}
\caption{Top: Binding energy vs. decay constant in three dimensions. Blue circles represent bound states induced by a local potential. From left to right, points correspond to increasing potential strength, with $g$ increasing from $4.5~t$ to $20~ t$ in steps of $0.5~t$. Red squares represent bound states induced by singularities of co-dimension three. From left to right, the degree of the singularity ($M$) increases in steps of $5$ from $5$ to $60$. The two data sets collapse onto the same curve. 
Bottom: $g_{M}$ vs. $M$, where $g_{M}$ is the potential that is equivalent to a singularity of degree $M$.   } 
\label{fig.3d_equivalence}
\end{figure}

We now compare results for a singularity with those for a potential. Bound states fit well to the same analytic form in both cases. As with the two-dimensional case, the decay constant and the binding energy are not independent. Fig.~\ref{fig.3d_equivalence} (top) plots the variation of these two parameters for singularity-driven and potential-driven bound states. Data from both cases collapse onto a single curve. This brings out a quantitative equivalence between singularities and potentials. For a singularity of degree $M$, we assign an equivalent potential, $g_M$. The degree-$M$ singularity produces a bound state with the same decay constant and binding energy as an attractive potential of strength $g_M$. Fig.~\ref{fig.3d_equivalence} (bottom) shows a plot of $g_M$ vs. $M$. For large $M$, $g_M$ approximately scales as $\sqrt{M}$. The equivalence is not restricted to the decay constant and binding energy. We have verified that it holds for the precise forms of the wavefunction, up to a change in the normalization constant.

\section{Discussion}

We have demonstrated that singularities arising from intersections produce bound states in the same way as attractive potentials. This mapping is quantitative in nature, where each singularity can be assigned an effective potential strength. In singularities, the binding mechanism is the kinetic energy of shuttling, where the particle moves back and forth across surfaces. This can be viewed as `quantum indecision' -- the particle remains frozen at a crossroads as it is unable to pick a direction of propagation. A bound state allows the system to sample all surfaces to small distances, while rapidly shuttling among surfaces. This notion can be tested in semiconductor architectures\cite{Bastard1988}, ultracold atomic gases\cite{Zakrzewski2007} and superconducting circuits\cite{Weiss2021}.

We have focussed on a class of spaces where the singularity is zero-dimensional, i.e., where smooth spaces intersect at a point. Within this class, the key factor that determines bound state formation is the dimensionality of the spaces involved. Our results can be generalized to higher-dimensional singularities with the key factor being co-dimensionality -- the difference in dimensionality between the smooth spaces and the singularity. For example, consider two 2D sheets that intersect along a line. In this case, we have translational symmetry along the intersection-line. This generates a new conserved quantity -- momentum along the intersection line. For each value of this momentum, we are left with an effective problem of two lines that intersect at a point. We conclude that a bound state can form for each momentum. All of these states may not be truly bound. If the kinetic energy of motion along the intersection-line exceeds the binding energy, this state can scatter and delocalize. An example of this physics is discussed in Sec.~IX of Ref.~\onlinecite{Khatua2019}.

Our results regarding the role of dimensionality in bound state formation are particularly relevant to quantum magnets. In the presence of frustration, the low energy physics of a magnet resembles a particle moving on an abstract space\cite{Khatua2019,khatua_thesis}. When this space self-intersects, the particle localizes. This manifests as magnetic ordering in a particular classical configuration. Co-dimension-1 singularities have been found and argued to host bound states\cite{Khatua2019,Sarvesh2020,Khatua2021}. Co-dimension-3 singularities have been found and argued \textit{not} to host bound states\cite{Khatua2018,khatua_thesis}. Building on these results, magnetic clusters can be designed to simulate spaces with multiple wires, sheets or even three-dimensional spaces. The Kitaev spin-$S$ chain serves as an example, with classical ground states that form a network-like space. Each node of the network is an intersection of $M$ wires, where $M$ can be tuned by changing the length of the chain\cite{Khatua2021}.

The analogy between singularities and potentials highlights the role of dimensionality in bound state formation. In higher dimensions, the tendency of a particle to spread is stronger as the space available for spreading is larger. As a result, a stronger potential or a singularity of higher degree is required. A similar idea is invoked in Anderson localization\cite{Economou1983,Economou1984}. In one or two dimensions, an infinitesimal amount of disorder suffices to localize a particle. However, a threshold disorder strength is required in three and higher dimensions. This has been related to the problem of a random walker and the mean time spent in a neighbourhood\cite{Berger2008}. The higher the dimensionality, the smaller is the time spent in a neighbourhood. The particle may show localizing tendencies, which upon quantization, manifest as bound states.

We have based our arguments on a tight binding framework where potentials and intersections can be handled on the same footing. 
More generally, the same problem can also be addressed in the continuum. There is extensive literature on quantum graphs where eigenfunctions of the Schr\"odinger operator can be found on each link, with suitable boundary conditions enforced at junctions. Studies have explored various choices for boundary conditions and their consequences\cite{Kottos1999,Znojil2012,Andrade2016}. Ref.~\onlinecite{Aharony2009} has compared the traditional quantum graph approach with tight binding (assuming plane-wave-like unbound states). With continuum problems on open/singular spaces, a careful self-adjoint formulation can give rise to bound states\cite{Tajron2022}. Our results pose an interesting question for future studies: what are the boundary conditions in the continuum problem that reproduce the tight binding bound state?

\acknowledgments
We thank Diptiman Sen, Kirill Samokhin, Jean-S\'ebastien Bernier, Subhankar Khatua and Abhiram Soori for insightful discussions. We thank Eric Tan for discussions on technical aspects. This work was supported by the Natural Sciences and Engineering Research Council of Canada.

\bibliographystyle{apsrev4-1} 
\bibliography{singularity}
\end{document}